\documentclass{SciPost}
\hypersetup{
    colorlinks,
    linkcolor={red!50!black},
    citecolor={blue!50!black},
    urlcolor={blue!80!black}
}

\usepackage[bitstream-charter]{mathdesign}
\usepackage{multirow}
\usepackage{bm}
\usepackage{physics}
\usepackage{amsmath}
\usepackage{tabularx}
\usepackage{booktabs} %

\definecolor{etapurple}{RGB}{164, 29, 173}
\DeclareSymbolFont{usualmathcal}{OMS}{cmsy}{m}{n}
\DeclareSymbolFontAlphabet{\mathcal}{usualmathcal}
\def \r{{\bm r}}
\def \q{{\bm q}}
\fancypagestyle{SPstyle}{
\fancyhf{}
\lhead{\colorbox{scipostblue}{\bf \color{white} ~SciPost Physics }}
\rhead{{\bf \color{scipostdeepblue} ~Submission }}

\fancyfoot[C]{\textbf{\thepage}}
}
\newcommand{\FigOne}{
\begin{figure}[t!]
\centering
\includegraphics[width=0.65\linewidth]{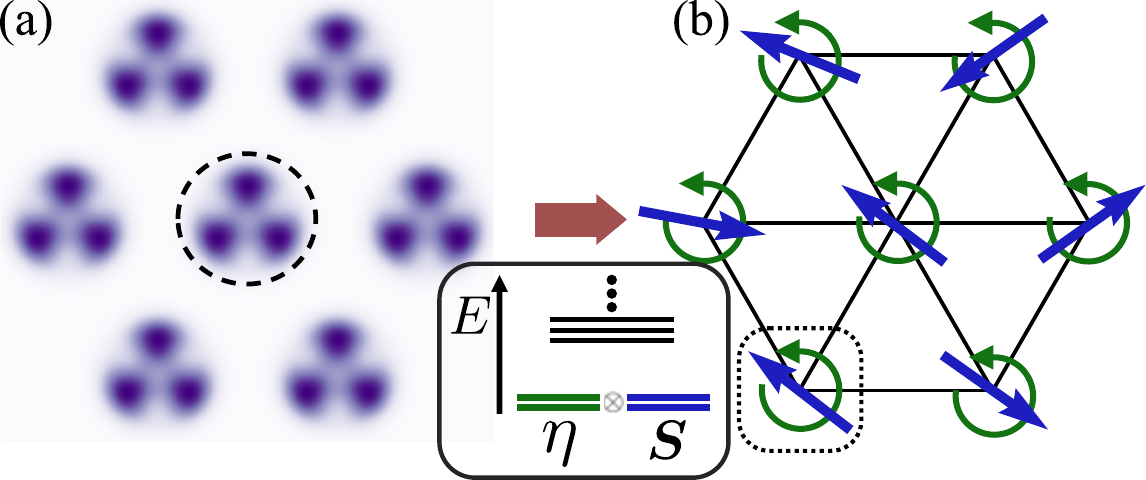}
\caption{(a) Charge density in a lattice of Wigner molecules. Each molecule (dotted circle) hosts three electrons localized at the vertices of an equilateral triangle, collectively forming a breathing kagome lattice. Inset: schematic energy spectrum a single molecule showing that the ground state manifold is composed of four degenerate levels: $(\eta^z,S^z) = (\pm 1, \pm 1/2)$. (b) Schematic of the local spin and orbital degrees of freedom interacting on a triangular lattice.}
\label{fig1}
\end{figure}
}
\newcommand{\FigTwo}{
\begin{figure}[t]
\centering
\includegraphics{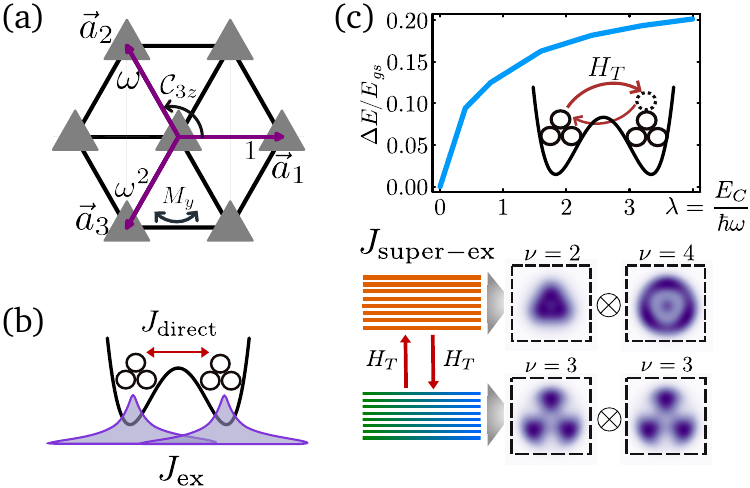}
\caption{
(a) The point group symmetries of $\mathcal{H}$ on the triangular lattice, with definitions of the lattice vectors $\bm{a}_i$, and bond-dependent phase factors $\nu_{ij}$. 
(b) Schematic of direct Coulomb repulsion $J_{\rm direct}$ between two Wigner molecules, and their exchange interaction $J_{\rm ex}$ from wave-function overlap.
(c) Super-exchange $J_{\rm super-ex}$ arising from virtual tunneling of electrons to excited states with $\nu \neq 3$.
Also shown are the charge gap $\Delta E(\lambda)$ (upper panel) and the drastic modification of Wigner molecular shapes at $\nu =2,4$ (lower panel). 
}
\label{fig2}
\end{figure}
}

\newcommand{\FigThree}{
\begin{figure}[!htbp]
\centering
\includegraphics{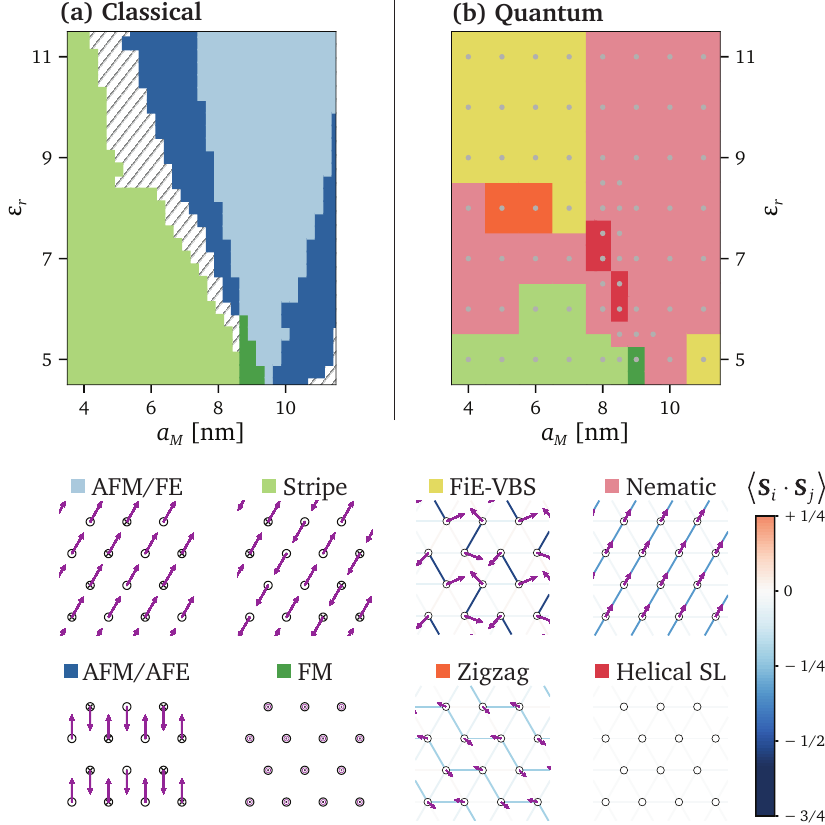}
\caption{(a) Phase diagram obtained by minimizing the energy over classical (pseudo)spin configurations at $(\phi, d_g) = (45^\circ, 20\,\mathrm{ nm})$. 
The classical phases are shown on the left; the Stripe and FM phases are also present in the quantum phase diagram. Purple arrows denote in-plane $\ev{\bm \eta_\parallel}$, purple circles (\textcolor{etapurple}{$\circ$}) denote positive $\ev{\eta^z}$, and black dots ($\odot$) and crosses ($\otimes$) denote positive and negative $\ev{S^z}$, respectively.
Our results are inconclusive in the cross-hatched region, where the minimum-energy states have large unit cells, suggesting high degeneracy or incommensurate order.
(b) Corresponding quantum phase diagram from DMRG, showing the emergence of new phases induced by quantum fluctuations. 
The in-plane orbital polarization $\langle \bm{\eta}_\parallel \rangle$ is shown by purple arrows, and spin-spin correlations on nearest neighbor bonds are indicated by color (blue represents antiferromagnetic correlations).  
Note that $\bm\eta$ does not transform like a vector under the reflection $M_y$, so the physical electrical polarization is given by $\hat{z} \times \langle \bm{\eta}_\| \rangle$ (see Table~\ref{tab:sym} and Fig.~\ref{fig5}). 
}
\label{fig3}
\end{figure}
}

\newcommand{\FigFour}{
\begin{figure}[t!]
\centering
\includegraphics{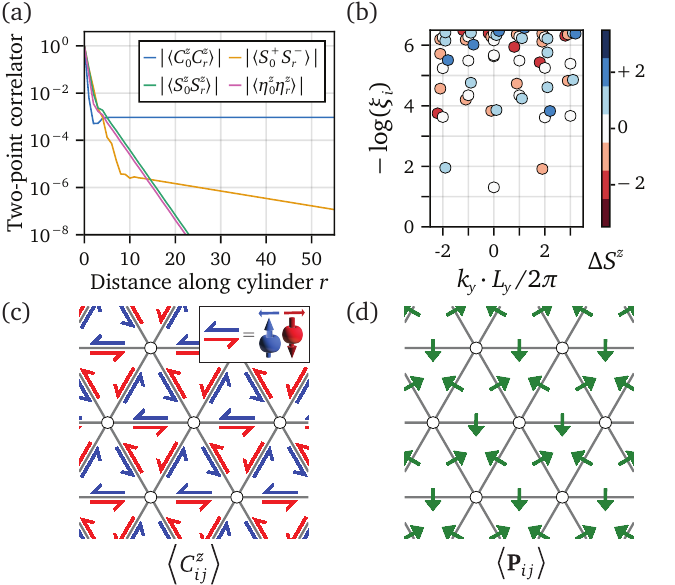}
\caption{Characterization of the HSL state found at $(a_M,\varepsilon_r) = (8.0 ~\mathrm{nm},7.0)$. 
(a) The exponential decay of the correlation functions $\ev{S^+_0 S^-_0}$, $\ev{\eta^z_0 \eta^z_r}$, and $\ev{S^z_0 S^z_r}$ indicates the lack of long-range order of either $\bm{S}$ or $\eta^z$.  
By contrast, $\langle C^z_0 C^z_r \rangle$ quickly approaches a constant plateau, signifying the ordering of the spin-currents. 
(b) The spin- and momentum-resolved entanglement spectrum exhibits two low-lying modes of opposite momentum and spin, indicating spin-momentum locking.
(c) Spin current pattern $C_{ij}$, which is of equal magnitude and sign on each bond. 
Since time-reversal is respected, along each bond, the spins of opposite orientation exhibit opposite currents of equal magnitude (blue and red arrows).
(d) The expectation value of vector orbital polarization $\bm{P}_{ij}$ (green arrows) is non-zero and $\mathcal{C}_{3v}$ symmetric.
} 
\label{fig4}
\end{figure}
}

\newcommand{\FigFive}{
\begin{figure}[t!]
\centering
\includegraphics[width=1.0\linewidth]{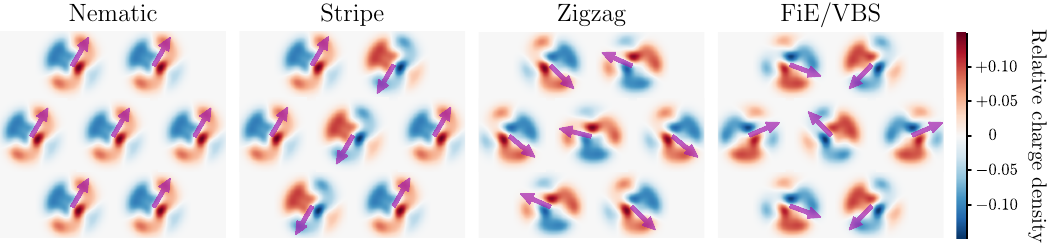}
\caption{
(a) Charge density distribution of the correlated phases that have local electrical polarization, i.e., $\hat{z} \times \langle \bm{\eta}_\parallel \rangle_i \neq 0$, subtracted from a state with no local electrical polarization ($\langle \bm{\eta}_\parallel \rangle_i = 0$) and normalized by the maximum density of this unpolarized state. Note that the electrical polarization is always rotated by 90$^\circ$ relative to the in-plane $\langle \bm{\eta}_\parallel \rangle_i$ indicated by purple arrows. Each phase has a distinct polarization pattern that serves as its unique fingerprint in STM.  
}
\label{fig5}
\end{figure}
}

\begin{document}

\pagestyle{SPstyle}

\begin{center}{\Large \textbf{\color{scipostdeepblue}{
Spin-orbital magnetism in moiré Wigner molecules\\
}}}\end{center}

\begin{center}\textbf{
Ahmed Khalifa\textsuperscript{1$\star$},
Rokas Veitas\textsuperscript{1},
Francisco Machado\textsuperscript{2,3,4}
,and
Shubhayu Chatterjee\textsuperscript{1$\dagger$}
}\end{center}

\begin{center}
{\bf 1} Department of Physics, Carnegie Mellon University, Pittsburgh, PA 15213, USA
\\
{\bf 2} ITAMP, Harvard-Smithsonian Center for Astrophysics, Cambridge, MA 02138, USA
\\
{\bf 3} Department of Physics, Harvard University, Cambridge, MA 02138, USA
\\
{\bf 4} QuTech, Delft University of Technology, PO Box 5046, 2600 GA Delft, The Netherlands 
\\[\baselineskip]
$\star$ \href{mailto:email1}{\small ahmedk@andrew.cmu.edu}\,,\quad
$\dagger$ \href{mailto:email2}{\small shubhayuchatterjee@cmu.edu }
\end{center}

\section*{\color{scipostdeepblue}{Abstract}}
\textbf{\boldmath{%
The interplay of spin and orbital degrees of freedom offers a versatile playground for the realization of a variety of correlated phases of matter.
    However, the types of spin-orbital interactions are often limited and challenging to tune.
    Here, we propose and analyze a new platform for spin-orbital interactions based upon a lattice of Wigner molecules in moir\'e transition metal dichalcogenides (TMDs). 
    Leveraging the spin-orbital degeneracy of the low-energy Hilbert space of each Wigner molecule, we demonstrate that TMD materials can host a general spin-orbital Hamiltonian that is tunable via the moir\'e superlattice spacing and dielectric environments. 
    We study the phase diagram for this model, revealing a rich landscape of phases driven by spin-orbital interactions, ranging from ferri-electric valence bond solids to a helical spin liquid.
    Our work establishes moir\'e Wigner molecules in TMD materials as a prominent platform for correlated spin-orbital phenomena.
}}

\vspace{\baselineskip}

\noindent\textcolor{white!90!black}{%
\fbox{\parbox{0.975\linewidth}{%
\textcolor{white!40!black}{\begin{tabular}{lr}%
  \begin{minipage}{0.6\textwidth}%
    {\small Copyright attribution to authors. \newline
    This work is a submission to SciPost Physics. \newline
    License information to appear upon publication. \newline
    Publication information to appear upon publication.}
  \end{minipage} & \begin{minipage}{0.4\textwidth}
    {\small Received Date \newline Accepted Date \newline Published Date}%
  \end{minipage}
\end{tabular}}
}}
}

\vspace{10pt}
\noindent\rule{\textwidth}{1pt}
\tableofcontents
\noindent\rule{\textwidth}{1pt}
\vspace{10pt}

\section{Introduction}
\label{sec:intro}
The complex interplay between orbital and spin degrees of freedom lies at the heart of some of the most exciting correlated many-body electronic phenomena~\cite{Kitaev_2006,JK2009,CJK2010,Singh2010,Comin2012,Li2015,Li2016,Luo2018,Kugel_1982,pavarini2008,Koga2018,Iwazaki2023,takagi2019concept}.
In the context of quantum magnetism,  spin-orbit coupling (SOC) is often responsible for the kinds of anisotropy and exchange frustrations that stabilize quantum spin liquid states~\cite{Kitaev_2006,JK2009,CJK2010}, while in spintronics, the coupling between electric and magnetic orders in multiferroics enables electrical control of magnetism~\cite{Matsukura2015,Ramesh2021,Sierra2021,DGG_2020_TBGswitch,Zhu2020,Polshyn2020,ABC2024}.
Observing this wealth of phenomena depends on our ability to control and engineer the underlying spin-orbit interactions.
However, in most materials, these interactions are usually dominated by a single term which ``locks'' the spin to the orbital degrees of freedom, such as momentum (Rashba SOC)~\cite{Manchon_2015,bihlmayer2022rashba,shi2023recent}, valley (Ising SOC)~\cite{zhang2023enhanced,zhang2025twist,patterson2025superconductivity,WVZC2024} and/or sublattice (Kane-Mele SOC)~\cite{KaneMele}.
Such locking suppresses quantum fluctuations and restricts our ability to explore more intricate phases of matter.

One tantalizing possibility for shaping spin-orbital interactions lies in the engineering of the orbital degrees of freedom.
Recent work in TMDs lays a promising path for such control~\cite{Reddy2023,PhysRevB.108.L121411,doi:10.1126/science.adk1348}.
More specifically, in bilayer moir\'e TMDs~\cite{andrei2021marvels,mak2022semiconductor,nuckolls2024microscopic}, the resulting superlattice potential can induce electronic flat bands with localized single-particle wavefunctions~\cite{wuHubbardModelPhysics2018,PhysRevResearch.2.033087,PhysRevB.102.201104}.
With one electron per moir\'e site, this setting realizes a spin-ful Hubbard model on the triangular superlattice ~\cite{wuHubbardModelPhysics2018,PhysRevResearch.2.033087,PhysRevB.102.201104,PhysRevB.102.235423,PhysRevB.107.235131,PhysRevB.109.045144}.
However, at a filling of three electrons (or holes) per unit cell, the strong interparticle Coulomb repulsion at each moir\'e site can split the electronic charge density into a localized trimer --- dubbed a Wigner molecule~\cite{Reddy2023,Luo2023,PhysRevB.108.L121411,PhysRevB.109.L121302,PhysRevLett.133.246502,li2024emergent,Li_2025,doi:10.1126/science.adk1348}.

\FigOne

In this work, we show that each Wigner molecule hosts a complex internal structure that leads to a rich family of spin-orbital Hamiltonians on the triangular lattice~[Fig.~\ref{fig1}(a)]. 
Our main results are threefold. 
First, we identify the low-energy Hilbert space for each Wigner molecule, via exact diagonalization (ED), as a degenerate four-dimensional space, charcterized by spin, $S = 1/2$, and orbital angular momentum, $L^z = \pm 1$, that acts as a pseudospin. 
Second, we derive an effective spin-orbital Hamiltonian constrained by symmetries and compute the couplings in this Hamiltonian using perturbation theory, demonstrating that the magnetic frustration and anisotropy can be tuned via moir\'e superlattice length-scale and dielectric screening.
Finally, using a combination of classical energy optimization and large-scale density matrix renormalization group (DMRG) calculation, we chart out the phase diagram of this Hamiltonian in the experimentally relevant regime. 
Specifically, we show that our spin-orbital model hosts a multitude of intriguing phases. 
These include not only classical multiferroic phases where orbital ferroelectricity intertwines with magnetic ordering of spins, but also intrinsically quantum phases such as a helical spin liquid characterized by spontaneous circulating spin-currents without any long-range magnetic order.
Collectively, our results demonstrate how TMD moiré materials offer a powerful and versatile platform for generating and studying a wealth of new spin-orbital phenomena.

\section{Wigner molecule local Hilbert space}
We begin by identifying the effective low-energy degrees of freedom of each Wigner molecule. 
To do so, we first review how Wigner molecules are formed in moir\'e TMD heterobilayers and $\Gamma$-valley  homobilayers~\cite{doi:10.1073/pnas.2021826118,Zhang2021,PhysRevLett.120.107703,PhysRevLett.131.046401,brzezinska2024pressure} (such as W$\mathrm{S}_2$).
In these materials, the low-energy single particle physics is described by the moir\'e potential $ V(\bm{r}) = -2V_{0} \left[ \sum_{\ell=1}^{3}\cos{(\bm{g}_{\ell}\cdot\bm{r}+\phi)} \right]$, where $\bm{g}_{\ell} = \frac{4\pi}{\sqrt{3}a_M}\left(\sin{\frac{2\pi \ell}{3}}, \cos{\frac{2\pi \ell}{3}}\right)$ are reciprocal moir\'e lattice vectors and $a_M$ is the moir\'e superlattice spacing.
The minima of $V(\bm{r})$ form a triangular lattice [Fig.~\ref{fig1}(a)].

To understand the structure of the moir\'e confining potential, we expand $V(\bm{r})$ on each site around its minimum such that
\begin{equation}
     V(\bm{r} - \bm{r}_{\min}) = V_{\mathcal{O}}(\bm{r}) \approx \frac{1}{2}kr^2 + c_3 \sin(3\theta) r^3~.
\label{eq:Vlocal}
\end{equation}
The confining potential is approximately harmonic (distorted by a weak trigonal moir\'e crystal field $c_3$) with an intrinsic oscillator frequency $\omega = \sqrt{k/m}$, where $k=16\pi^2V_0\cos\phi / a_M^2$  and $m$ is the band mass of the electron.
Consequently, each moir\'e superlattice site can host localized electrons (or holes), with a confinement length-scale $\xi_0 = (\hbar^2/mk)^{1/4} \lesssim a_M$.

This single-particle description is strongly modified by interactions;
intra-moir\'e-site Coulomb repulsion imbues the system with a charge gap at integer filling $\nu$, while at higher fillings [$\nu\ge 2$], it modifies the structure of the localized wavefunctions.
To this end, let us consider the following Hamiltonian for each moiré site:
\begin{equation}
    \mathcal{H}_{\text{site}} = \sum_{i=1}^{\nu}\left[\frac{\bm{p}_i^2}{2m} + V_{\mathcal{O}}(\bm{r}_i)\right] + \sum_{i<j}\mathcal{V}(\bm{r}_i - \bm{r}_j),
\label{eq:WMH} 
\end{equation}
where $\mathcal{V}(\bm{r}_i - \bm{r}_j) = \frac{e^2}{\varepsilon}\left(\frac{1}{\abs{\bm{r}_{i}-\bm{r}_{j}}}-\frac{1}{\sqrt{\abs{\bm{r}_i - \bm{r}_j}^2+4d_g^2}}\right)$ is the single-gate screened Coulomb interaction. 
$d_g$ denotes the distance to the metal gate and $\varepsilon = \varepsilon_0 \varepsilon_r$ quantifies the electric screening due to the dielectric substrate. 
The presence of strong Coulomb interactions within each site repels the electrons apart, reshaping the local charge density and forming a Wigner molecule~\cite{PhysRevLett.82.5325,PhysRevLett.82.3320,Mikhailov:2002,PhysRevLett.96.126806,Kalliakos2008}, the zero-dimensional equivalent of a Wigner crystal.
In contrast to the conventional $\nu=1$ Mott insulator~\cite{wuHubbardModelPhysics2018,AuerbachBook}, the Wigner molecule can host additional orbital degrees of freedom, whose fluctuations enables novel quantum phases.

We focus on $\nu = 3$, where the local charge density takes a `three-lobed' trimer form within a single molecule \cite{Reddy2023, PhysRevB.108.L121411}, resulting in a breathing Kagome lattice [Fig.~\ref{fig1}(a)] stabilized by the trigonal moir\'e crystal field. 
The simplest way to understand the low energy manifold of the Wigner molecule is by focusing on the non-interacting harmonic limit.
In this limit, the single-particle energy eigenstates are given by $E(n,l^z,s^z) = (2 n + |l^z| + 1) \hbar \omega$ \cite{Mikhailov:2002}, where each level has a two-fold degeneracy due to SU(2)$_s$ spin-rotation symmetry of the spin-1/2 electron. 
The ground state manifold is then easy to obtain: the $(n, l^z, s^z) = (0,0,\pm 1/2)$ states are filled with two electrons, while the third electron has four degenerate options, $(n,l^z,s^z) = (0,\pm 1, \pm 1/2)$.
These states have a total angular momentum of $L^z = \sum_{i=1}^{3} l^z_i = \pm 1$ and total spin of $S^z = \sum_{i=1}^{3} s^z_i = \pm 1/2$. 
This picture is quite robust.
First, the trigonal warping [$c_3 \neq 0$ in Eq.~\ref{eq:Vlocal}] reduces the spatial symmetry from U(1) to $\mathcal{C}_{3v}$ so that $L^z$ is only defined modulo 3, yet $L^z = \pm 1$ remain good quantum numbers.
Second, the combination of time-reversal $\cal{T}$ and SU(2)$_s$ symmetry constrain eigenstates with $L^z = \pm 1$ and $S^z = \pm 1/2$ to remain degenerate \emph{even when strong electron-electron interactions significantly re-shape the charge distribution.}
Therefore, below a critical interaction strength, the effective local Hilbert space remains exactly four-dimensional.

We confirm our intuitive expectation with an explicit ED computation of the spectrum of $\mathcal{H}_{\text{site}}$ \cite{Reddy2023, PhysRevB.108.L121411}. 
Parameterizing the intra-site interaction strength by the ratio of the intra-unit-cell Coulomb repulsion, $E_c = e^2/\varepsilon \xi_0$, to the harmonic level spacing, $\lambda = E_c/\hbar \omega$, we find that for $\lambda < \lambda_c \approx 4.5$, the low-energy state-space remains four-dimensional. 
This space can be characterized as a tensor product of two spin-like degrees of freedom: 
(i) the total spin of the electrons, represented by $\bm{S} = (S^x, S^y, S^z)$, 
and (ii) a pseudospin-half degree of freedom capturing the total angular momentum $L^z = \pm 1$, represented by $\bm{\eta} = (\eta^x, \eta^y, \eta^z)$.
Because the angular momentum is restricted to only two levels, the explicit mapping between $\bm{L}$ and $\bm{\eta}$ is given by $L^z\rightarrow\eta^z$, $\frac{1}{2}(L^+)^2\rightarrow\eta^+$, and $\frac{1}{2}(L^-)^2\rightarrow\eta^-$.
Crucially, this mapping shows that, while $\eta^z$ remains odd under $\cal{T}$, the in-plane components of the orbital-pseudospin $\bm{\eta}$, namely $\bm{\eta}_\parallel = (\eta^x, \eta^y)$ are \textit{even} under $\cal{T}$ and transform as a pseudovector under $\mathcal{C}_{3v}$ (see Appendix~\ref{app:pol} for a detailed symmetry analysis).
This implies that the $\hat{z} \times \bm{\eta}_\parallel$ behaves as an electrical polarization that couples linearly to in-plane electric field $\bm{E}_\parallel$, and ordering of $\bm{\eta}_\parallel$ leads to ferroelectricity.

To conclude this section, we note that for large intra-molecular Coulomb repulsion $\lambda > \lambda_c$,  Hund's rule prevails and the interacting ground state has higher total spin $S = 3/2$ and $L^z = 0$~\cite{Thouless_1965}. 
In this limit, SU(2)$_s$ symmetry constrains the triangular Wigner molecular lattice to realize a $S = 3/2$ Heisenberg model, whose ground state is either the ferromagnet or the 120$^\circ$ antiferromagnet \cite{Olariu2009}. 
Therefore, in this work, we focus on $\lambda < \lambda_c$, where the spin and orbital degrees of freedom enable a plethora of frustrated magnetic and multiferroic phases, and turn to deriving the spin-orbital Hamiltonian that describes interacting Wigner molecules on a lattice.

\begin{table}[h!]
\begin{center}
\begin{tabular}{ |c|| c|c| c|} 
 \hline
 \multirow{2}{*}{Symmetry}&  \multirow{2}{*}{Spin ($\bm{S}_i$) } & \multicolumn{2}{c|}{Pseudospin}\\
 &  & (${\eta}^+_i$) &  ($\eta^z_i$)
\\
\hline \hline 
 \rule{0pt}{3ex}$\mathcal{C}_{3z}$ & $U_s^\dagger\left(\hat{z}, \frac{2\pi}{3}  \right)  \bm{S}_{\mathcal{C}_{3z} \bm{r}_i} U_s\left(\hat{z}, \frac{2\pi}{3}  \right)$ & $\omega^2 \eta^+_{\mathcal{C}_{3z} \bm{r}_i}$ & $\eta^z_{\mathcal{C}_{3z} \bm{r}_i}$ \\ 
 \rule{0pt}{4ex}$M_y$ & $(S^x_{M_y \bm{r}_i}, -S^y_{M_y \bm{r}_i}, -S^z_{M_y \bm{r}_i})$ & $ \eta^-_{M_y \bm{r}_i}$ & $- \eta^z_{M_y \bm{r}_i}$  \\ 
 \rule{0pt}{4ex}$\mathcal{T}$ & $-\bm{S}_i$ & $\eta^{-}_i$ &  $- \eta^z_i$ \\
 \rule{0pt}{4ex}SU(2)$_s$ & $U_s^\dagger\left(\hat{\bm{n}}, \theta  \right)  \bm{S}_i  U_s\left(\hat{\bm{n}}, \theta  \right)$ & $\eta^+_i$ & $\eta^z_i$
 \\
 \hline 
\end{tabular}

\caption{Point-group and internal symmetry actions on the low-energy degrees of freedom $(\bm{S}, \bm{\eta})$ of Wigner molecules at $\nu = 3$. 
The three columns show the result of each symmetry transformation on the operators $\bm{S}_i$, $\eta^+_i$, and $\eta^z_i$, respectively. Here, $U(\hat{\bm{n}},\theta) = e^{-i \theta \hat{\bm{n}} \cdot \bm{S}/\hbar }$ indicates SU(2)$_s$ spin-rotation about $\hat{\bm{n}}$ by angle $\theta$, $\mathcal{C}_{3z} \bm{r}_i$ is the rotation of the vector $\bm{r}_i$ by $2\pi/3$ about the $z$-axis, $M_y \bm{r_i} = M_y(x_i, y_i) \equiv (-x_i,y_i)$ is a spatial reflection, and $\omega \equiv e^{2\pi i/3}$ is a cube root of unity.}
\label{tab:sym}
\end{center}
\renewcommand{\arraystretch}{1}
\end{table}

\section{Effective spin-orbital model}
\label{sec:another}
Having understood the low-energy manifold of each Wigner molecule, we can leverage our understanding of the system's symmetries to constrain the form of the effective spin-orbital Hamiltonian. 
The system exhibits a $\mathcal{C}_{3v}$ point group symmetry generated by a three-fold rotation $\mathcal{C}_{3z}$ and mirror $M_y$ about the $yz$-plane [Fig.~\ref{fig2}(a)], as well as time-reversal $\cal{T}$ and SU(2)$_s$ spin-rotation symmetry. 
We list the symmetry actions on the $(\bm{S},\bm{\eta})$ operators in Table~\ref{tab:sym}.
Whereas spin-rotation symmetry constrains the spin-spin interactions to be Heisenberg-like, the reduced rotational symmetry of the orbital pseudospin $\bm{\eta}$ allows for terms that either couple to the spin-Heisenberg term or appear as bilinears of $\bm{\eta}$ operators only.
Taken together, the most general symmetry-constrained Hamiltonian, focusing on nearest neighbor interactions, is given by (see \ref{app:Eff-H} for more details): 
\begin{align}
\label{eq:spin_Ham}
    \notag
   \mathcal{H} &= \sum_{\langle ij\rangle} \bigl[\left(J_H + H_{zz}^{s} + H_{+-}^{s} + H_{++}^{s} + H_{+}\right) \bm{S}_{i}\cdot\bm{S}_{j}\\ 
   & \hspace{10mm}+\left(H_{zz}^o + H_{+-}^o + H_{++}^o\right)\bigr],\\[2mm]
   H_{zz}^{\alpha} &= J_{zz}^{\alpha}\eta^z_i\eta^z_j \notag \\ \notag
   H_{+-}^{\alpha} &= J_{+-}^{\alpha}e^{i\theta_{+-}^\alpha}\eta^+_i \eta^-_j + h.c.\\\notag
   H_{++}^{\alpha} &= J_{++}^{\alpha}\left(\nu_{ij}\eta^+_i \eta^+_j + h.c.\right)\\\notag
   H_{+} &= J_{+}e^{i\theta_{+}}\left(\nu_{ij}^\ast\eta^+_i +\nu_{ij}\eta^-_j\right) + h.c.\notag
\end{align}
where $\{ J^\alpha \}$ denote the different couplings, superscript $\alpha$ takes value $s$ ($o$) for spin-orbital (orbital only) terms, and $\nu_{ij} = 1, \omega, \omega^2$ for bonds along lattice vectors $\bm{a}_1, \bm{a}_2, \bm{a}_3$, respectively. 
The general form of $\mathcal{H}$ puts forth an expressive model that captures a wide variety of correlated spin-orbital physics.

    \FigTwo

Crucially, the expressibility of $\mathcal{H}$ can be accessed in a wide range of  materials, notably heterobilayer and homobilayer TMD stacks such as twisted bilayer WS$_2$~\cite{doi:10.1126/science.adk1348}, and MoS$_2$~\cite{Naik2020}, heterobilayer WSe$_2$/WS$_2$ and twisted double-bilayer WSe$_2$~\cite{Foutty2023}.
Given the tunability of these platforms, both in terms of the choice of van der Waals materials, but also twist angle, applied fields, and dielectric substrates, this expressibility can translate itself into the experimental exploration of a wealth of intricate orbital-magnetic phenomena.
To highlight some of these features, we focus on the particular case of twisted bilayer WS$_2$, where a Wigner molecule lattice has been experimentally observed~\cite{doi:10.1126/science.adk1348};
we now proceed to compute the various couplings for this material platform and study its phase diagram.

\FigThree

The different couplings $J^\alpha$ are determined by three distinct physical effects: direct Coulomb repulsion, quantum mechanical exchange, and tunneling mediated super-exchange~\cite{AuerbachBook}.
The direct Coulomb interaction naturally couples the orbital degrees of freedom; the asymmetric charge density profile at each site generates an electric field which directly couples to the charge density at nearby sites.
Owing to the long-range nature of the Coulomb interaction, this effect is significant when the distance between neighboring sites is smaller than the gate-screening length, $a_M \lesssim d_g$.
By contrast, the exchange interaction arises from the overlap of different Wigner molecules. 
Therefore, it is most significant for small $a_M$ and $\varepsilon_r$, where each Wigner molecule is larger.
As $a_M$ or $\varepsilon_r$ increase, super-exchange interactions, arising from virtual tunneling of electrons between adjacent Wigner molecules [Fig.~\ref{fig2}(c)], start to dominate over exchange processes.
In this regime, spin-spin interactions become antiferromagnetic, and the system becomes frustrated.
Despite this simple description, the nature of the (super-)exchange interaction is qualitatively modified by the orbital degrees of freedom.
Unlike the conventional Hubbard model~\cite{AuerbachBook}, the shape of each Wigner molecule is drastically modified as the number of electrons at each site changes [Fig.~\ref{fig2}(d)], inducing more complex matrix elements and orbital-spin interactions.
This subtle interplay manifests itself into a rich phase diagram.

\section{Landscape of spin-orbital phenomena} 
We begin by considering classical phases captured by symmetry breaking patterns in $\langle \bm{S} \rangle$ and/or $\langle \bm{\eta}\rangle$.
To this end, we constrain $\langle \bm{\eta}\rangle$ and $2\langle \bm{S}\rangle$ to the unit sphere, and minimize $\mathcal{H}[\langle \bm{S} \rangle,\langle \bm{\eta}\rangle]$ via manifold conjugate gradient descent~\cite{kofod2025,edelman1998,absil2008}, while considering enlarged unit cells to capture different translation symmetry breaking patterns~\cite{LUS}.
We fix $(\phi, d_g) = (45^\circ, 20~\mathrm{nm})$, typical values for WS$_2$ ~\cite{doi:10.1126/science.adk1348}, and determine the minimum energy configuration of $\langle \bm{S}\rangle,\langle \bm{\eta}\rangle$ as a function of $(a_M,\varepsilon_r)$, which can be directly tuned by stacking and twisting of the TMD materials, and their dielectric environment, respectively.
The resulting phase diagram is summarized in Fig.~\ref{fig3}(a).

The simplest phase to understand is the spin-orbital ferromagnet (FM), which breaks both time-reversal $\mathcal{T}$ and SU(2)$_s$ spin rotation symmetries. 
This phase occurs when both the spin coupling $J_H$ and the spin-orbital coupling $J^s_{zz}$ are strongly ferromagnetic.
The system is no longer highly frustrated, and both spin $\bm{S}$ and orbital magnetic moment $\eta^z$ can exhibit homogeneous, non-zero expectation values.
This offers a different path towards the realization of spin-orbital ferromagnetism that, unlike anomalous Hall phases in moir\'e TMDs  \cite{PhysRevLett.122.086402,LiQAH2021,PhysRevX.13.031037,PhysRevX.14.011004,Cai2023}, does not require the engineering of flat bands with significant Berry curvature.

When $a_M$ is reduced, $\cal{H}$ becomes dominated by the $J_+$ term which linearly couples the in-plane orbital pseudospin $\bm{\eta}_\parallel$ to the Heisenberg spin-bilinear on bonds [Eq.~\ref{eq:spin_Ham}]. 
In position space, the energy of a particular classical configuration can then be understood as follows. 
The pattern of spin-correlations on the bonds $\langle ij \rangle$ generates a local field for $ \bm{\eta}_{\parallel}$ at site $i$; the energy of the state is dictated by the strength of the field which determines the orientation of  $\bm{\eta}_{\parallel,i}$.
Owing to the $\mathcal{C}_{3v}$ symmetry of $\cal{H}$, the local field is exactly canceled when the system exhibits a ferromagnetic spin-ordering with uniform $\langle \bm{S}_i \cdot \bm{S}_j \rangle$---therefore, the lowest energy configuration must exhibit a non-trivial pattern of spin correlations.
Indeed, we observe the formation of a \emph{stripe} phase where both $\langle \bm{\eta}_\parallel \rangle$ and $\langle \bm{S} \rangle$ display an alternating collinear pattern, albeit with different periodicities, breaking both translation and rotational symmetries~[Fig.~\ref{fig3}(a)].
Specifically, $\langle \bm{\eta}_\parallel \rangle_i \sim \bm{a}_3 \cos(2\bm{Q}_s \cdot \bm{r}_i)$ oscillates twice as fast as the spin density $\langle \bm{S} \rangle_i \sim \hat{z} \left[\cos(\bm{Q}_s \cdot \bm{r}_i) + \sin(\bm{Q}_s \cdot \bm{r}_i)\right]$, with an ordering wave-vector $\bm{Q}_s = \pi(\hat{z} \times \bm{a}_3)/(\sqrt{3} a_M^2)$ that is perpendicular to $\langle \bm{\eta}_\parallel \rangle$. 
Both the period-doubling and the relative orientation of $\langle \bm{\eta}_\parallel \rangle$ and $\bm{Q}_s$ can be explained using the Landau theory of stripes~\cite{ZKS98} --- the lowest-order symmetry-allowed coupling between the local charge density $\rho_b$ and spin-density $\langle \bm{S} \rangle$ takes the form $\rho_b \langle \bm{S} \rangle^2$. 
Accordingly, spin-ordering at momentum $\bm{Q}_s$ leads to a charge density oscillating at $2\bm{Q}_s$, corresponding to a longitudinal electrical polarization at the same wave-vector. 
This is indeed borne out by our data: 
the electrical polarization $\hat{z} \times \langle \bm{\eta}_\parallel \rangle_i \sim \bm{Q}_s \cos(2\bm{Q}_s \cdot \bm{r}_i)$ precisely encodes a charge density~\cite{boundcharge}
that oscillates twice as fast as the spin-density.
Lastly, we note that the coexistence of antiferroelectricity and magnetism implies that this stripe phase is multiferroic, suggesting a new avenue for designing materials with spintronic applications~\cite{shi2023recent}.

The rest of the classical phase diagram is either populated by an antiferromagnet, coexisting with in-plane ferroelectric (AFM/FE) or antiferroelectric (AFM/AFE) ordering (blue regions in Fig.~\ref{fig3}(a)), or by ever more complex patterns of magnetism, characterized by large unit cells [hatched region in Fig.~\ref{fig3}(a), see Appendix~\ref{app:class} for details].
The prevalence of several competing magnetic configurations in the classical phase diagram is a strong indication that quantum fluctuations play an important role in understanding the zero temperature phase diagram of $\mathcal{H}$.

To this end, we employ large-scale cylinder-iDMRG simulations with circumference $L_y = 4-6$ to find the quantum ground state of $\cal{H}$~\cite{tenpy2024}.
The resulting phase diagram can be divided into three broad categories~[Fig.~\ref{fig3}(b)]. 
The first category are the phases already captured by the classical phase diagram;
it includes both the spin-orbital ferromagnet and the stripe order, whose classical and quantum energy densities are nearly identical [Appendix~\ref{app:class}].

The second category includes point group symmetry breaking phases that are stabilized by quantum correlations [Fig.~\ref{fig3}(b)].
These correlations give rise to either a Valence Bond Solid (VBS), where pairs of sites form spin singlets, or phases characterized by decoupled spin chains that realize one-dimensional Luttinger liquid physics~\cite{giamarchi2003quantum}.

The VBS phase emerges whenever $\mathcal{H}$ is dominated by frustration-inducing antiferromagnetic interactions.
Indeed, the location of this phase matches with the existence of competing classical configurations with large unit cells; quantum fluctuations drive the system to a VBS state with a non-collinear in-plane ordering for $\bm{\eta}_\parallel$ over a four-site unit cell.
We dub this phase a ferri-electric valence bond solid (FiE-VBS) phase, as it features an in-plane electrical polarization owing to the non-zero net value of $\langle \bm{\eta}_\parallel \rangle$.

The alternative pattern of $\mathcal{C}_{3v}$ symmetry breaking occurs when the spin degrees of freedom form one dimensional channels rather than singlets.
This can either occur along a principal lattice direction [``Nematic'' phase in Fig.~\ref{fig3}(b)], or along more intricate paths [``Zigzag'' phase in Fig.~\ref{fig3}(b)].
These patterns of antiferromangetic correlations occur on top of classical ordering of $\bm{\eta}_{\|}$.
This observation offers a simple picture for origin of these states: the local orbital polarization $
\langle \bm{\eta}_{\|} \rangle_i$ modulates the spin-spin interaction along different bonds and enhances the antiferromagnetic coupling along certain paths, thereby inducing strong 1D magnetic correlations in a 2D system~\cite{Zheng2006,Hayashi2007,Heidarian2009}.

\FigFour

The final category exhibits no onsite order and preserves all point group symmetries---this is the hallmark of a spin liquid state.
In the parameter regime exhibiting the greatest competition between different phases, we observe a state characterized by exponentially decaying correlations of single-site spin and orbital operators [Fig.~\ref{fig4}(a)].
The point group symmetric nature is clear when considering single site and two-site \emph{spin} operators, but when focusing on two-body orbital correlations, it requires additional care.
A naive observation of the $\langle \eta^x_i\eta^x_j \rangle$ would suggest that the point group symmetry is broken.
However, because this operator does not transform either as a scalar or a vector under $\mathcal C_{3v}$, it is not an accurate indicator of $\mathcal C_{3v}$ breaking. 
To this end, we construct the orbital bilinear operator $\bm{P}_{ij} = (\eta^x_i\eta^y_j + \eta^y_i\eta^x_j, \eta^x_i\eta^x_j - \eta^y_i\eta^y_j)$ that transforms as a vector [Appendix~\ref{app:pol}]; indeed, $\langle \bm{P}_{ij} \rangle$ clearly respects the symmetry of the model~[Fig.~\ref{fig4}(d)].

Remarkably, although the system lacks onsite order and preserves both $\mathcal{C}_{3v}$ and translation symmetries, it breaks the SU(2)$_s$ spin-rotation symmetry.
This is manifest in the spin-current operator $\bm{C}_{ij} = \bm{S}_i\times \bm{S}_j$; in Fig.~\ref{fig4}(c), we consider $C^z_{ij}$ on nearest neighbor bonds $\langle ij \rangle$, showing that it exhibits long-range order~[Fig.~\ref{fig4}(a)]~\cite{PinningField}.
This ordering pattern highlights an intriguing feature of this phase: even though SU(2)$_s$ spin-rotation symmetry is broken, time reversal symmetry is preserved --- for this reason we term this phase the Helical Spin Liquid (HSL).

We conclude with a few remarks about the HSL phase.
First, the HSL phase features non-zero expectation values for both bond-polarization $\bm{P}_{ij}$ and spin-current $\bm{C}_{ij}$. 
This is reminiscent of a magnetoelectric effect that arises in strongly spin-orbit coupled non-collinear spiral magnets, where a spin-current can linearly couple to the bond-polarization~\cite{Katsura2005}, implying that these observables either both vanish, or are both non-zero.
In sharp contrast, the presence of SU(2)$_s$ spin-rotation symmetry forbids such a linear coupling between $\bm{P}_{ij}$ and $\bm{C}_{ij}$ in the Wigner-molecular crystal. 
As such, the bond polarization $\bm{P}_{ij}$ is generically allowed by symmetry and exists in most of the phase diagram, while the spin-current $\bm{C}_{ij}$ breaks SU(2)$_s$ and is non-zero only in the HSL phase.  
Second, although we only observe long-range ordering for the spin-current, the spin correlation $\langle S^+_0 S^-_r \rangle$ exhibits a correlation length much larger than the other local correlation functions.
While the exact source remains unclear, the phase of the correlator matches a periodicity of three sites, suggesting a significant susceptibility to a 120$^\circ$ ordering that decreases with increasing bond dimension.
Finally, the HSL exhibits a spontaneous, \emph{interaction-driven} generation of spin-momentum locking in a SU(2)$_s$ symmetric setting.
This is in contrast to previous proposals for such locking (in time-reversal and SU(2)$_s$ symmetric systems), where one either has a mean-field fermionic band picture describing a topological insulator~\cite{Raghu2008}, or higher-body, ring-exchange type interactions~\cite{Troyer2005}.
This locking is most clearly observed by computing the half-system entanglement spectrum~\cite{li:2008a, qi:2012a}; the low-lying spectrum exhibits two degenerate modes with opposite momentum and opposite spin [Fig.~\ref{fig4}(b), blue and red].

\section{Experimental signatures}

\FigFive

Having studied the wealth of phenomena supported in this platform, we now turn to a discussion of the requirements and opportunities for their experimental characterization.

To diagnose the presence of magnetic order, scanning SQUID measurements or reflective magnetic circular dichroism (RMCD) already serve powerful probes of moir\'e heterostructures~\cite{tschirhart2021imaging,anderson2023programming,XY2025}; the same tools can be used to explore the orbital ferromagnetism in the spin-orbital ferromagnetic state in Wigner molecules.
By contrast, deciphering the nature of correlated phases which lack a net magnetic moment, such as spin/orbital antiferromagnets or quantum spin liquids, has proven challenging in moir\'e materials~\cite{KIVC,KIVC2,IKS}. 
The presence of an additional orbital pseudospin degree of freedom $\bm{\eta}$ in Wigner molecules offers new routes to overcome this challenge. 
More specifically, because the in-plane component $\bm{\eta}_\parallel$ is a time-reversal even pseudovector and $\hat{z} \times \bm{\eta}_\parallel$ behaves akin to an electrical polarization, one can more easily address and detect the textures of $\bm{\eta}_\parallel$ and probe the underlying ordered phase.
Indeed, both static and dynamical signatures of the electrical polarization can be used to delineate a large fraction of the phase diagram.

At the static end, spontaneous electrical polarization leads to a re-arrangement of the charge density distribution, which can be directly visualized via scanning tunneling microscopy (STM).  
Specifically, different $\langle \bm{\eta}_\parallel \rangle$ patterns map onto distinct charge densities, which can be studied relative to a state with no electrical polarization (e.g., the spin-orbital ferromagnet).
This acts as a fingerprint for the nematic, stripe, zigzag and ferrielectric-VBS phases [Fig.~\ref{fig5}].
Unfortunately, the HSL state features no on-site polarization or magnetization; however, as the only studied state without any form of either order, the observation a featureless phase flanked by other phases with detectable electric/magnetic order will strongly hint towards the presence of the HSL.

To further disambiaguate these states, we can probe their dynamical properties.
Each correlated phase displays a distinct symmetry breaking pattern, determining a nature and spectrum of the low-energy quasi-particles.
These characteristics are encoded in the electromagnetic fluctuations nearby the material which can be studied using proximate spin-qubits~\cite{Joaquin2018,CRD2019,sahay2021noise,machado2023quantum,ziffer2024quantum}, such as nitrogen-vacancy (NV) centers in diamond~\cite{rovny2024nanoscale}.

While the spin-orbital correlated states in Wigner molecules are electrically insulating at $T = 0$, many of the broken-symmetries are expected to be stable to thermal fluctuations due to their discrete nature or finite meso-scale sample sizes. 
As such, one may still perform low-temperature transport measurements and observe hysteresis loops with in-plane electric fields or out-of-plane magnetic fields to detect the presence of ferroelectricity or ferromagnetism.
Further, the low-lying states in the entanglement spectrum of the HSL [Fig.~\ref{fig4}(b)] indicate possible low-energy edge modes with spin-momentum locking that may be characterized via edge transport.

Finally, all-optical probes, which can avoid regions of twist angle inhomogeneities via concentrated laser spots, may also be used to probe the underlying charge and spin-density distribution via a modification of excitonic and polaronic resonances~\cite{sidler2017fermi,ImamogluPRL2019}. 
This is particularly relevant to heterobilayer moir\'e TMDs, where the formation of correlated Wigner molecular phases in one layer can directly affect the excitonic spectrum in the other layer. 
The quantum twisting microscope~\cite{inbar2023quantum}, which is able to probe momentum-resolved features by performing local interference experiments with its tip, may also be used determine the order-parameter momenta for translation-symmetry broken phases. 

\section{Conclusion}
Our work proposes moir\'e TMD materials as a novel platform for spin-orbital correlated matter.
We demonstrate that such materials can naturally host lattices of Wigner molecules, and feature enhanced quantum fluctuations that lead to intriguing multiferroic and spin liquid phases. 
In particular, we find an elusive time-reversal symmetric helical spin liquid, which features spontaneous spin-currents, in a microscopically realizable spin-orbital Hamiltonian with only nearest neighbor interactions. 

An important question to consider with regards to the experimental feasibility of observing such spin-orbital correlated phases is their stability to thermal fluctuations. 
On generic grounds, we expect that phases that break discrete symmetries, i.e., any phase with $\bm{\eta}$-ordering, is stable to a temperature scale of tens of Kelvin, corresponding to the typical coupling strength $J$ of a few meVs: the critical temperature for each phase may be determined by numerical Monte-Carlo studies. 
We note that the Mermin-Wagner-Hohenberg theorem~\cite{MerminWagner,Hohenberg} prohibits the breaking of the continuous spin-rotation symmetry, e.g., as expected in the helical liquid phase, at any non-zero temperature in the thermodynamic limit.
Nevertheless, the order-parameter correlation length grows exponentially at low temperatures $k_B T \ll J$ for such phases, making their experimental observation feasible in meso-scale devices. 

The tunability of the Wigner molecular platform extends beyond our studied spin-orbital model, suggesting a wide range of future explorations.
First, so far our work was restricted to investigating the impact of the moiré lengthscale and electric permittivity.
The plethora of tuning parameters available in such TMD materials (including strain, stacking order and configuration, electric and magnetic fields, etc.) greatly extends the parameter ranges accessible in the model of Eq.~\eqref{eq:spin_Ham}.
At the same time, the nature of the localized degrees of freedom in our Wigner molecule platform offers new paths for controlling the underlying order.
In particular, because $\bm{\eta_\|}$ is time-reversal even, it couples linearly to electric fields but not to magnetic fields, allowing the independent use of in-plane electric fields to polarize $\bm{\eta}$ and in-plane magnetic fields to polarize spin $\bm{S}$.
Second, while we focused on $\Gamma$-valley TMDs in this work, our proposal can naturally be adapted to $K$-valley TMDs which feature spin-valley locking, reducing SU(2)$_s$ to U(1)$_s$~\cite{xiao2012coupled,mak2022semiconductor}.
Finally, lattices of Wigner molecules can also be realized using bosonic degrees of freedom (such as moir\'e excitons~\cite{QXtoDX2024}), enabling the study of role of quantum statistics in these correlated settings.

More broadly, our proposal motivates future studies of TMDs in a less explored regime, requiring new tools to study and probe spin-orbital correlations. 
This could have important implications for realizing unconventional phases of quantum matter and developing novel materials for spintronic applications.

\section*{Acknowledgements}
We gratefully acknowledge helpful discussions with M.~Bintz, S.~Chern, M.~Crommie, E.~Davis, S.~Divic, E.~Khalaf, A.~N.~Pasupathy, J.~Sau, N.~Verma, F.~Wang, T.~A.~Webb and M.~P.~Zaletel. This work used Bridges-2 at PSC through allocation PHY240241 from the Advanced Cyberinfrastructure Coordination Ecosystem: Services \& Support (ACCESS) program~\cite{access}, which is supported by U.S. National Science Foundation grants No. 2138259, No. 2138286, No. 2138307, No. 2137603, and No. 2138296.
F.~M.~acknowledge support from the NSF through a grant for
ITAMP at Harvard University.

\begin{appendix}
\numberwithin{equation}{section}

\section{Exact diagonalization on a single moir\'e site}
For the sake of completeness, we review the 2D harmonic oscillator which we use to form our single-particle basis in order to obtain the spectrum of $N$ interacting electrons. The expansion of the moir\'e potential up to order $r^6$ and ignoring the constant shift is:
\begin{equation}
    \begin{aligned}
        V_{\mathcal{O}}(\bm{r}) &= 8V_0\pi^2\cos(\phi)\left(\frac{r}{a_M}\right)^2+\frac{16V_0\pi^3}{3\sqrt{3}}\sin(\phi)\sin(3\theta)\left(\frac{r}{a_M}\right)^3-\frac{8V_0\pi^4}{3}\cos(\phi)\left(\frac{r}{a_M}\right)^4\\
        &-\frac{16V_0\pi^5}{9\sqrt{3}}\sin(\phi)\sin(3\theta)\left(\frac{r}{a_M}\right)^5+\frac{16V_0\pi^6}{405}\cos(\phi)(10-\cos(6\theta))\left(\frac{r}{a_M}\right)^6. 
    \end{aligned}
\end{equation}
Truncating to the second order, we get the harmonic oscillator problem with the Hamiltonian,
\begin{equation}
    H = \frac{p^2}{2m} + \frac{1}{2}kr^2,
\end{equation}
where $k = 16V_0\pi^2\cos(\phi)/a_M^2$. In order to make use of the U(1) symmetry of the Hamiltonian, we define creation and annihilation operators that make the symmetry manifest,
\begin{equation}
    a_{\pm} = \frac{1}{2}\left[\frac{x\mp iy}{\xi_0}+\frac{\xi_0}{\hbar}(p_x \mp ip_y)\right],
\end{equation}
where we defined $\xi_0\equiv\left(\hbar^2/mk\right)^{1/4}$ as the characteristic length scale of the harmonic oscillator. 
In terms of the creation and annihilation operators, the Hamiltonian becomes
\begin{equation}
    H = \hbar\omega(a_{+}^{\dagger}a_{+}+a_{-}^{\dagger}a_{-}+1).
\end{equation}
The energy spectrum is simply $E = \hbar\omega\left(n_+ + n_- + 1\right)$, where $n_{\pm} \in \{0,1,2,\ldots\}$, and the eigenstates are,
\begin{equation}
     \ket{n_+n_-} = \frac{(a_+^{\dag})^{n_+}(a_-^{\dag})^{n_-}}{\sqrt{n_+!n_-!}}\ket{00}
\end{equation}
These states are also eigenstates of the $z$-component of orbital angular momentum operator, $L^z = xp_y - yp_x = a_{+}^{\dagger}a_{+} - a_{-}^{\dagger}a_{-}$ with eigenvalues $l^z = n_+ - n_-$. So an equivalent label of the states is given by $n, l^z$ where $n\equiv min(n_+, n_-)$. The real space wave functions are the Fock-Darwin orbitals, and they are given by
\begin{equation}
    \psi_{n,l^z,s^z}(\bm{r})=\frac{(-1)^{n}}{\xi_0}\sqrt{\frac{n!}{\pi(n+|l^z|)!}} \, e^{il^z\theta} \, \left(\frac{r}{\xi_0}\right)^{|l^z|}L_{n}^{|l^z|}\left(\frac{r^2}{\xi_0^2}\right) \, e^{-\frac{r^2}{2 \xi_0^2}}\chi_{s^z},
\end{equation}
where $s^z = \pm 1/2$ for up/down spin. 

In our exact diagonalization (ED) calculations, we use these single-particle orbitals to form a basis made of single Slater determinants (SSD) for the states of $N$ electrons. so that a general many-body wave function is a linear combination of these states as,
\begin{equation}
    \Phi_m\left(\bm{r}_1,\ldots,\bm{r}_N\right) = \sum_i c_i^m \Psi_{SSD}^i\left(\bm{r}_1,\ldots,\bm{r}_N\right)
\end{equation}
We then write the Hamiltonian operator as a matrix which we diagonalize to get the ground state and the first few excited states. 
If we define $n\equiv n_+ + n_- +1$ as a collective label for the single-particle orbitals, we keep up to $n_{max} = 8$ as a cutoff which translates to 36 spin orbitals. The number of many-body states in the basis set is $1028790$, $59640$, $2556$ for $N=4$, $N=3$, $N=2$, respectively. We check the convergence by calculating the percent error in the ground state energy as the basis cutoff is changed. We find $(E_{28}-E_{36})/E_{28}\approx0.1\%$ which indicates that the basis cutoff of 36 spin orbitals is justified.
We also make use of the symmetry of the system which makes the Hamiltonian block-diagonal in the total $S^z$ and $L^z ~\mathrm{mod} ~3$.

\begin{figure}
    \centering
    \includegraphics[width=0.5\linewidth]{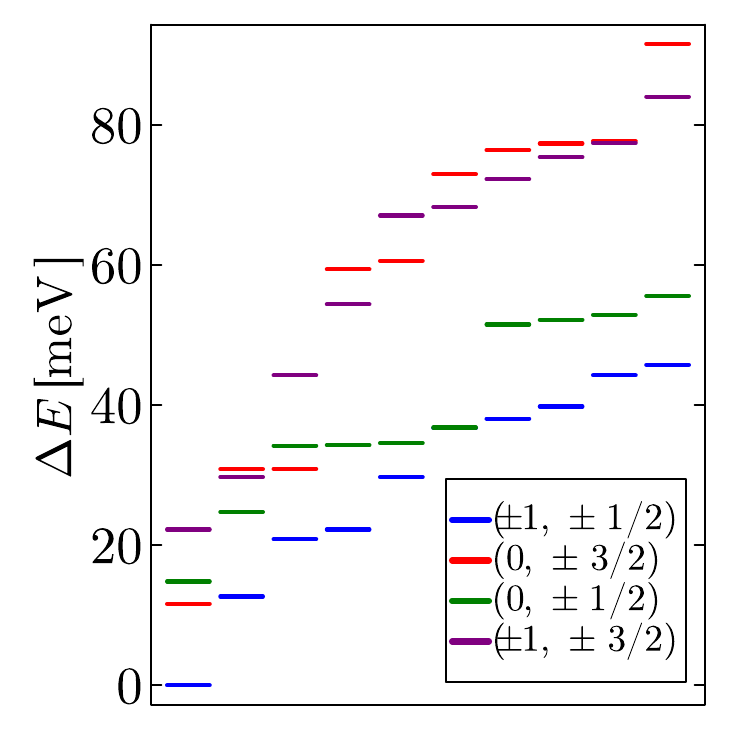}
    \caption{The ED spectrum of three interacting electrons confined around the moir\'e potential minimum. the ground state is in the sector $(L^z = \pm1, S^z = \pm1/2)$ separated by a gap to excited states.}
    \label{fig:ED_site}
\end{figure}

Finally, our ED calculation on a single site supports the validity of projecting onto the four fold degenerate manifold of states which is the basis of our spin-orbital model. The validity is based on two factors. First, the ground state has to be in the sector with total S = 1/2 and total $L_z = \pm 1$. second, the ground state is separated by a gap from excited states. This is controlled by the ratio of the Coulomb energy scale to the harmonic confinement energy scale $\lambda = E_C/\hbar\omega$ where for $\lambda < 4.5$ the low spin state is the ground state. To relate this ratio to our microscopic parameters, we note that as $\varepsilon_r$ becomes smaller, $E_C$ gets bigger and hence $\lambda$ increases. Similarly $\hbar\omega \propto 1/a_M$ so as $a_M$ gets bigger $\lambda$ increases. So by choosing $\varepsilon_r$ and $a_M$ appropriately we can be in a regime where these two factors are satisfied. A typical spectrum in this parameters regime is shown in Figure \ref{fig:ED_site} where the gap between the $(L^z = \pm1, S^z = \pm1/2)$ and the first excited state is of the order $15-20$ meV.  

\section{Coulomb matrix elements}
\subsection{Direct term}
Here, we present the derivation of the Coulomb matrix elements which are decomposed into direct and exchange terms. We closely follow the method of \cite{Reddy2023}, and we generalize it to the general situation for our purposes which includes intersite interaction with gate screening. 
The Coulomb repulsion between electrons at neighboring sites on the triangular lattice separated by a distance $a_M$ along the x-axis (say the origin and $a_M \hat{x}$), is given by
\begin{equation}
    U(\r + a_M \hat{x}) = \frac{2\pi e^2}{\epsilon}\int \frac{d^{2}q}{(2\pi)^2}U(\q)e^{i\q\cdot\r}e^{iq_x a_M}
\end{equation}

where we use the single gate screened Coulomb interaction 
\begin{equation}
U(\q) = \frac{1-e^{-2qd}}{q}    
\end{equation}

Henceforth, we will implicitly measure distances in units of the harmonic confinement length-scale $\xi_0 = (\hbar^2/mk)^{1/4}$, and momenta in units of $\xi_0^{-1}$.
Accordingly, we define the complex momentum (measured in units of $\xi_0^{-1}$) as 
\begin{equation}
    \Tilde{q} = \frac{q_x + iq_y}{2}
\end{equation}
The Coulomb matrix element is 
\begin{equation}
   U_{ijkl} = \int \frac{d^2 q}{(2\pi)^2}\frac{1-e^{-2qd}}{q}e^{iqa_M\cos{\phi}}\bra{n_{+i}n_{-i},n_{+j}n_{-j}}e^{i\q\cdot(\r_1 - \r_2)}\ket{n_{+k}n_{-k},n_{+l}n_{-l}}
\end{equation}
In terms of $\tilde{q}$, we have 
\begin{multline}
    U_{ijkl} = 2E_C \int \frac{d^2 \tilde{q}}{2\pi}U(\tilde{q},\phi) \bra{n_{+i}}e^{i(\tilde{q}a_+^\dagger+\tilde{q}^\ast a_+)}\ket{n_{+k}} \bra{n_{-i}}e^{i(\tilde{q}a_- +\tilde{q}^\ast a_-^\dagger)}\ket{n_{-k}}\\
   \bra{n_{+j}}e^{-i(\tilde{q}a_+^\dagger+\tilde{q}^\ast a_+)}\ket{n_{+l}}\bra{n_{-j}}e^{-i(\tilde{q}a_- +\tilde{q}^\ast a_-^\dagger)}\ket{n_{-l}}
\end{multline}

where $U(\tilde{q},\phi) = \frac{1-e^{-4|\tilde{q}|d}}{|\tilde{q}|} e^{2i |\tilde{q}|a_M\cos{\phi}}$. The matrix element becomes

\begin{align}
    \begin{split}
        U_{ijkl} &= 2E_c(-1)^{N_j+N_k}i^{\Delta N}(n_{i+}!n_{j+}!n_{k+}!n_{l+}!n_{i-}!n_{j-}!n_{k-}!n_{l-}!)^{1/2}\int\,\frac{d^2\tilde{q}}{2\pi}U(\tilde{q},\phi)e^{-2\tilde{q}^2}e^{i\Delta l \phi} \\
        &\times \left(\sum_{\alpha=0}^{\text{min}(n_{i+},n_{k+})}\frac{(-1)^{\alpha}|\tilde{q}|^{n_{i+}+n_{k+}-2\alpha}}{\alpha!(n_{i+}-\alpha)!(n_{k+}-\alpha)!}\right)\left(\sum_{\beta=0}^{\text{min}(n_{i-},n_{k-})}\frac{(-1)^{\beta}|\tilde{q}|^{n_{i-}+n_{k-}-2\beta}}{\beta!(n_{i-}-\beta)!(n_{k-}-\beta)!}\right) \\
        &\times \left(\sum_{\gamma=0}^{\text{min}(n_{j+},n_{l+})}\frac{(-1)^{\gamma}|\tilde{q}|^{n_{j+}+n_{l+}-2\gamma}}{\gamma!(n_{j+}-\gamma)!(n_{l+}-\gamma)!}\right)\left(\sum_{\eta=0}^{\text{min}(n_{j-},n_{l-})}\frac{(-1)^{\eta}|\tilde{q}|^{n_{j-}+n_{l-}-2\eta}}{\eta!(n_{j-}-\eta)!(n_{l-}-\eta)!}\right) \\
        &= 2E_c(-1)^{N_j+N_k}i^{\Delta N}(n_{i+}!n_{j+}!n_{k+}!n_{l+}!n_{i-}!n_{j-}!n_{k-}!n_{l-}!)^{1/2}\\
        &\times \sum_{\alpha=0}^{\text{min}(n_{i+},n_{k+})}\frac{(-1)^{\alpha}}{\alpha!(n_{i+}-\alpha)!(n_{k+}-\alpha)!} \sum_{\beta=0}^{\text{min}(n_{i-},n_{k-})}\frac{(-1)^{\beta}}{\beta!(n_{i-}-\beta)!(n_{k-}-\beta)!} \\
        &\times \sum_{\gamma=0}^{\text{min}(n_{j+},n_{l+})}\frac{(-1)^{\gamma}}{\gamma!(n_{j+}-\gamma)!(n_{l+}-\gamma)!}\sum_{\eta=0}^{\text{min}(n_{j-},n_{l-})}\frac{(-1)^{\eta}}{\eta!(n_{j-}-\eta)!(n_{l-}-\eta)!}I_q(p)
    \end{split}
\end{align}

\begin{equation}
    I_q(p) = \int \frac{d^{2}\tilde{q}}{2\pi}e^{-2\tilde{q}^2}\tilde{q}^{p-1}(1-e^{-4\tilde{q}d})e^{i\Delta l \phi}e^{2i\tilde{q}a_M\cos{\phi}},
\end{equation}
where $\Delta l = (n_{i+}-n_{i-} + n_{j+}-n_{j-}) - (n_{k+}-n_{k-} + n_{l+}-n_{l-})$,\\ 
and $p = (n_{i+}+n_{k+}-2\alpha)+(n_{i-}+n_{k-}-2\beta)+(n_{j+}+n_{l+}-2\gamma)+(n_{j-}+n_{l-}-2\eta)$, and we have defined $\tilde{q} \equiv |\tilde{q}|$ in $I_q(p)$ to simplify notations henceforth.

A special case for the above expression is  $a_M=0$, which corresponds to the intra-site Coulomb interaction with screening. The integral becomes 
\begin{equation}
    I_q(p)\big|_{a_M=0} = \int_0^\infty d\tilde{q}e^{-2\tilde{q}^2}\tilde{q}^{p}(1-e^{-4\tilde{q}d}) = 2^{\frac{-3}{2}}2^{\frac{-p}{2}}\left[\Gamma\left(\frac{p+1}{2}\right)-\Gamma(p+1)U\left(\frac{p+1}{2},\frac{1}{2},2d^2\right)\right],
\end{equation}
where $U$ is the confluent hypergeometric function.

Another special case is the case of inter-site interaction but with no screening, i.e., $d = 0$. We can do the $\phi$ integral first to get 
\begin{equation}
    I_\phi = \frac{1}{2\pi}\int_0^{2\pi} d\phi e^{i\Delta l\phi} e^{2i\tilde{q}a_M \cos{\phi}} = i^{\Delta l}J_{\Delta l}(2\tilde{q}a_M)~~~~\mathrm{for}~\Delta l > 0
\end{equation}
For $\Delta l < 0$, we have $I_\phi = i^{3\Delta l} J_{\abs{\Delta l}}(2\tilde{q}a_M)$. 
Then we have 
\begin{multline}
    I_q(p)\big|_{d=0} = i^{\Delta l sgn(\Delta l)}\int_0^\infty d\tilde{q}  e^{-2\tilde{q}^2}\tilde{q}^{p}J_{\abs{\Delta l}}(2\tilde{q}a_M) = i^{\Delta l sgn(\Delta l)}2^{\frac{-3}{2}}2^{-\frac{p+\abs{\Delta l}}{2}}\frac{\Gamma\left(\frac{p+\abs{\Delta l} +1}{2}\right)}{\Gamma(\abs{\Delta l} +1)}a_M^{\abs{\Delta l}}\\
    M\left(\frac{p+\abs{\Delta l} +1}{2},\abs{\Delta l}+1,-\frac{a_M^2}{2}\right)
\end{multline}
Both the above integrals can be found in \cite{gradshteyn2007}.

The most general case does not have a closed form. So the remaining integral is when $d\neq 0,~a_M\neq 0$.

\begin{equation}
    I_{d,a_M} = -\int \frac{d^{2}\tilde{q}}{2\pi}e^{-2\tilde{q}^2}\tilde{q}^{p-1}e^{-4\tilde{q}d}e^{i\Delta l \phi}e^{2i\tilde{q}a_M \cos{\phi}}
\end{equation}

First we do the radial integral \begin{equation}
    I_\phi = \int_0^\infty d\tilde{q}e^{-2\tilde{q}^2}\tilde{q}^{p}e^{-4\tilde{q}d}e^{2i\tilde{q}a_M\cos{\phi}} = 2^{\frac{-3}{2}}2^{-\frac{3p}{2}}\Gamma(p+1)U\left(\frac{1+p}{2},\frac{1}{2},\frac{(2d-ia_M\cos{\phi})^2}{2}\right)
\end{equation}

Then $I_{d,a_M}$ becomes 
\begin{equation}
    I_{d,a_M} =  -2^{\frac{-3}{2}}2^{-\frac{3p}{2}}\Gamma(p+1)\int_0^{2\pi}\frac{d\phi}{2\pi}e^{i\Delta l \phi}U\left(\frac{1+p}{2},\frac{1}{2},\frac{(2d-ia_M\cos{\phi})^2}{2}\right)
\end{equation}
which we evaluate numerically to extract the direct Coulomb term. 

\subsection{Exchange term}

Our previous calculations only capture the direct part of the intersite Coulomb interactions. Now we will extend it. First, recall this is the Coulomb interaction operator  
\begin{equation}
    U(\r_1 - \r_2) = \frac{e^2}{\epsilon}\int \frac{d^{2}q}{2\pi}U(\q)e^{i\q\cdot(\r_1 - \r_2)}
\end{equation}

We will compute the matrix element of this operator in the basis $(n_x, n_y)$ of the harmonic oscillator which takes the form,

\begin{equation}
   U_{ijkl} = E_C \int \frac{d^2 q}{2\pi}U(\q)\int d^{2}r_1 d^{2}r_2 \psi_{n_{xi}, n_{yi}}(\r_1)\psi_{n_{xj}, n_{yj}}(\r_2 - a_M \hat{x} )e^{i\q\cdot(\r_1 - \r_2)}\psi_{n_{xk}, n_{yk}}(\r_1 - a_M \hat{x})\psi_{n_{xl}, n_{yl}}(\r_2)
\end{equation}
Note that we dropped the complex conjugate because the basis functions are real. The central object of this calculation is the form factor $F(\q)$, given by 
\begin{equation}
    \begin{split}
    F_1(\q) &= \int d^{2}r_1  \psi_{n_{xi}, n_{yi}}(\r_1)e^{i\q\cdot\r_1}\psi_{n_{xk}, n_{yk}}(\r_1 - a_M \hat{x})\\ 
    &= \int_{-\infty}^{\infty} dx  \psi_{n_{xi}}(x)e^{iq_{x}x}\psi_{n_{xk}}(x - a_M)
    \int_{-\infty}^{\infty} dy  \psi_{n_{yi}}(y)e^{iq_{y}y}\psi_{n_{yk}}(y)
    \end{split}
\end{equation}
As in the previous section, we measure the position (momentum) in units of $\xi_0$ ($1/\xi_0$). 
The $y$ integral can be evaluated simply per Ref.~\cite{lopez2000matrix}
\begin{equation}
    \bra{m}e^{-\gamma y}\ket{n} = \sqrt{2^{n-m}\frac{n!}{m!}}(-\gamma)^{m-n}e^\frac{\gamma^2}{4}L_n^{m-n}\left(\frac{-\gamma^2}{2}\right)
\end{equation}
We may assume $m > n$ without loss of generality as the matrix element is symmetric. So in order to apply this formula to our case, we have $m_1 = max(n_{yi}, n_{yk})$ and $m_2 = max(n_{yl}, n_{yj})$ We also assume without any loss of generality that $n_{yi}>n_{yk}$ and $n_{yj}>n_{yl}$. This leads to the following for the $y$ integral,

\begin{equation}
    I_y = \sqrt{2^{n_{yk}-n_{yi}}\frac{n_{yk}!}{n_{yi}!}}(iq_y)^{n_{yi}-n_{yk}}e^{-\frac{q_y^2}{4}}L_{n_{yk}}^{n_{yi}-n_{yk}}\left(\frac{q_y^2}{2}\right)
\end{equation}

Where the associated Laguerre polynomial is,

\begin{equation}
    L_n^{q}(x) = \sum_{k=0}^{n} \frac{(-1)^k (n+q)! x^k}{k! (k+q)! (n-k)!}
\end{equation}

We can then write $I_y$ as,
\begin{equation}
    I_y = \sqrt{2^{n_{yk}-n_{yi}}n_{yk}!n_{yi}!}(i)^{n_{yi}-n_{yk}}e^{-\frac{q_y^2}{4}}\sum_{p=0}^{n_{yk}}\frac{(-1)^pq_y^{2p+n_{yi}-n_{yk}}}{2^pp!(p+n_{yi}-n_{yk})!(n_{yk}-p)!}
\end{equation}
For the $x$ integral we use the result in Ref.~\cite{GUSEINOV2006226}
\begin{equation}
   I_x = \frac{(-1)^{n_{xi}+n_{xk}}e^{\frac{iq_x a_M}{2}}e^{\frac{-q_x^2-a_M^2}{4}}}{\sqrt{2^{n_{xi}+n_{xk}}n_{xi}!n_{xk}!}} \sum_{i=0}^{n_{xi}}\sum_{j=0}^{n_{xk}} \sum_{k=0}^{\lfloor\frac{i+j}{2}\rfloor} \binom{n_{ix}}{i}\binom{n_{xk}}{j} H_{n_{xi}-i}\left(\frac{-a_M}{2}\right)H_{n_{xk}-j}\left(\frac{a_M}{2}\right) \frac{(i+j)!(-iq_x)^{i+j-2k}}{(i+j-2k)!k!},
\end{equation}
 where $H_n (x)$ is a Hermite polynomial. 

The $\r_2$ integral is similar with the replacements of $i\q\rightarrow -i\q$, $n_{xi}\rightarrow n_{xl}$, $n_{xk}\rightarrow n_{xj}$.

Now we are left with the $\q$ integral. We collect all the terms depending on $\q$, 

\begin{equation}
    I_q = \int\frac{d^2 q}{2\pi}U(\q)e^{-q^2/2}(-iq_x)^{n_1-2k_1}(iq_x)^{n_2-2k_2}q_y^{2p_1+2p_2+n_{yi}+n_{yj}-n_{yk}-n_{yl}}
\end{equation}
then the only terms depending on $\q$ gives,
\begin{equation}
\begin{split}
    I_q &= \int\frac{d^2 q}{2\pi}U(\q)e^{-q^2/2}q_x^{n_1+n_2-2k_1-2k_2}q_y^{2p_1+2p_2+n_{yi}+n_{yj}-n_{yk}-n_{yl}}\\
         &= \frac{1}{2\pi}\int_0^{2\pi}d\phi\int_0^\infty dq U(q)e^{-q^2/2}q^{n_1+n_2-2k_1-2k_2+2p_1+2p_2+n_{yi}+n_{yj}-n_{yk}-n_{yl}}\\
         & (\cos{\phi})^{n_1+n_2-2k_1-2k_2}(\sin{\phi})^{2p_1+2p_2+n_{yi}+n_{yj}-n_{yk}-n_{yl}}, 
\end{split}
\end{equation}
where $n_1 = i_1 +j_1$ and $n_2 = i_2 +j_2$.

Note that this integral and the matrix element vanish if $n_{yi}+n_{yj}-n_{yk}-n_{yl}$ is odd. Let $p = 2p_1+2p_2+n_{yi}+n_{yj}-n_{yk}-n_{yl}+n_1+n_2-2k_1-2k_2$, $p_x = n_1+n_2-2k_1-2k_2$, $p_y = 2p_1+2p_2+n_{yi}+n_{yj}-n_{yk}-n_{yl}$. With that in mind, the result of the integral becomes,
\begin{equation}
    I_q = \frac{(1+(-1)^{p_x})^2 \Gamma\left(\frac{p_x+1}{2}\right)\Gamma\left(\frac{p_y+1}{2}\right)}{4\pi\Gamma\left(\frac{2+p}{2}\right)}2^{\frac{-p}{2}}\left[2^{\frac{-1}{2}+p}\Gamma\left(\frac{p+1}{2}\right)-d \Gamma(p+1)U\left(p+\frac{1}{2},\frac{3}{2},2d^2\right)\right]
\end{equation}

Finally, we can write down the result for the matrix element. 
\begin{equation}
\begin{aligned}
    U_{ijkl} = \mathcal{N}\sum_{i_1, i_2, j_1, j_2}\sum_{k_1, k_2, p_1, p_2}&\bigg[\binom{n_{xi}}{i_1}\binom{n_{xl}}{i_2}\binom{n_{xk}}{j_1}\binom{n_{xj}}{j_2}H_{n_{xi}-i_1}\left(\frac{-a_M}{2}\right)\\
    &\times H_{n_{xl}-i_2}\left(\frac{-a_M}{2}\right)H_{n_{xk}-j_1}\left(\frac{a_M}{2}\right)H_{n_{xj}-j_2}\left(\frac{a_M}{2}\right) \\
    &\times \frac{(-i)^{i_1 + j_1 -2k_1}(i_1 + j_1)!}{(i_1 + j_1 -2k_1)!k_1 !}\frac{i^{i_2 + j_2 -2k_2}(i_2 + j_2)!}{(i_2 + j_2 -2k_2)!k_2 !}\\
    &\times \frac{(-1)^{p_1}}{2^{p_1}p_{1}!(p_1+n_{yi}-n_{yk})!(n_{yk}-p_1)!}\\
    &\times \left.\frac{(-1)^{p_2}}{2^{p_2}p_{2}!(p_2+n_{yj}-n_{yl})!(n_{yl}-p_2)!}I_{q}(k_1, k_2, p_1, p_2) \right],
\end{aligned}
\end{equation}
$\mathcal{N} = \frac{E_C}{4\pi}\frac{i^{n_{y1}}(-i)^{n_{y2}}\sqrt{2^{-N_y}n_{yk}!n_{yi}!n_{yj}!n_{yl}!}}{\sqrt{2^{N_x}n_{xk}!n_{xi}!n_{xj}!n_{xl}!}}e^{-\frac{a_M^2}{2}}$, where $n_{y1} \equiv max(n_{yi}, n_{yk}) - min(n_{yi}, n_{yk})$ and $n_{y2} \equiv max(n_{yj}, n_{yl}) - min(n_{yj}, n_{yl})$.

\section{The effective spin-orbital Hamiltonian}\label{app:Eff-H}

We go into the details of how the various $J^\alpha$ couplings in the effective spin model (Eq.~(3) of the main text) are determined using perturbation theory. 
We consider two nearest neighbor sites in the triangular lattice $(i, j)$ that are related to each other by reflection; the remaining couplings can be obtained from symmetry. 
The Hilbert space of the degenerate ground state manifold is 16-dimensional. It is spanned by the superposition of states $\ket{\Psi_{is\eta}}\otimes\ket{\Psi_{js^\prime\eta^\prime}}$ that forms an $S=0$ singlet and an $S=1$ triplet. Due to the SU(2)$_s$ symmetry, the Hamiltonian is block diagonal in this basis.  We denote the block of $S=0$ by $\mathcal{M}_s$ and the $S=1$ block by $\mathcal{M}_t$, referring to singlet and triplet, respectively. 

The singlet sector includes the states,
\begin{equation}
    \begin{aligned}
        \ket{\Psi_1^s} &= \frac{1}{\sqrt{2}}\left(\ket{-1,\uparrow}_i\otimes\ket{1,\downarrow}_j-\ket{-1,\downarrow}_i\otimes\ket{1,\uparrow}_j\right)\\
        \ket{\Psi_2^s} &= \frac{1}{\sqrt{2}}\left(\ket{1,\uparrow}_i\otimes\ket{-1,\downarrow}_j-\ket{1,\downarrow}_i\otimes\ket{-1,\uparrow}_j\right)\\
        \ket{\Psi_3^s} &= \frac{1}{\sqrt{2}}\left(\ket{-1,\uparrow}_i\otimes\ket{-1,\downarrow}_j-\ket{-1,\downarrow}_i\otimes\ket{-1,\uparrow}_j\right)\\
        \ket{\Psi_4^s} &= \frac{1}{\sqrt{2}}\left(\ket{1,\uparrow}_i\otimes\ket{1,\downarrow}_j-\ket{1,\downarrow}_i\otimes\ket{1,\uparrow}_j\right),
    \end{aligned}
\end{equation}
where $\pm1$ denotes the orbital degrees of freedom and the arrows denote the spin. Similarly, we have for the triplet sector with $S^z = 1$ the states,
\begin{equation}
    \begin{aligned}
        \ket{\Psi_1^t} &= \ket{-1,\uparrow}_i\otimes\ket{1,\uparrow}_j\\
        \ket{\Psi_2^t} &= \ket{1,\uparrow}_i\otimes\ket{-1,\uparrow}_j\\
        \ket{\Psi_3^t} &= \ket{-1,\uparrow}_i\otimes\ket{-1,\uparrow}_j\\
        \ket{\Psi_4^t} &= \ket{1,\uparrow}_i\otimes\ket{1,\uparrow}_j
    \end{aligned}
\end{equation}

We start with the inter-site Coulomb interaction denoted by $V_C$ which contributes to first order in perturbation theory.
We calculate the matrix elements of the inter-site Coulomb interaction (using the expressions derived in the previous sections) between states in the ground state manifold $n,m$, $\mathcal{M}^{(1)}_{mn} = \bra{\Psi_m}V_{C}\ket{\Psi_n}$. 

\begin{figure}[t!]
\centering
\includegraphics{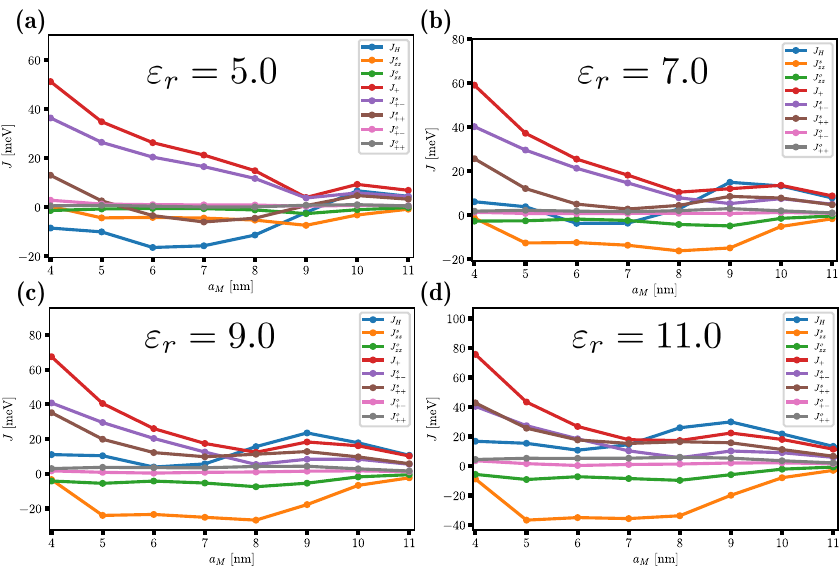}
\caption{(a)-(d)The dependence of the couplings $J$ as the moir\'e scale $a_M$ is varied at four different values of $\varepsilon_r$.} 
\label{fig:J_aM}
\end{figure}

Next, we consider the effect of the moir\'e potential, which hops an electron through to a neighboring site and back. This is a second-order process, so we need second order degenerate perturbation theory to get $\mathcal{M}^{(2)}_{mn}$,
\begin{equation}
    \mathcal{M}^{(2)}_{mn} = \sum_{\rm ex}\frac{\bra{m}H_{T}\ket{\rm ex}\bra{\rm ex}H_{T}\ket{n}}{E_{0} - E_{\rm ex}}
\end{equation}
where $\ket{\rm ex}$ are the intermediate excited states where the two sites hold $2$ and $4$ electrons respectively.

We can parametrize the two matrices $\mathcal{M}_\alpha$ by,
\begin{equation}
    \mathcal{M}_{\alpha} = \begin{pmatrix} \mathcal{M}_\alpha^{11} & \mathcal{M}_\alpha^{12} & \mathcal{M}_\alpha^{13} & \mathcal{M}_\alpha^{13}\\
    {\mathcal{M}_\alpha^{12}}^\ast & \mathcal{M}_\alpha^{11} & {\mathcal{M}_\alpha^{13}}^{\ast} & {\mathcal{M}_\alpha^{13}}^{\ast}\\
    {\mathcal{M}_\alpha^{13}}^\ast & \mathcal{M}_\alpha^{13} & \mathcal{M}_\alpha^{33} & \mathcal{M}_\alpha^{34}\\
    {\mathcal{M}_\alpha^{13}}^\ast & \mathcal{M}_\alpha^{13} & \mathcal{M}_\alpha^{34} & \mathcal{M}_\alpha^{33}
    \end{pmatrix},
\end{equation}
where $\alpha$ denotes $s,t$. A general 16 by 16 matrix is spanned by 256 tensor product of Pauli matrices and the identity matrix. We set $\mathcal{M}$ equal to a linear combination of these matrices,
\begin{equation}
    \label{eq:Mdecomp}
    \mathcal{M} = \sum_{\alpha\beta\gamma\delta}J_{\alpha\beta\gamma\delta}S^{\alpha}_i\otimes\eta^{\beta}_i\otimes S^{\gamma}_j\otimes\eta^{\delta}_j.
\end{equation}
The solution to the set of linear equations obtained from Equation (\ref{eq:Mdecomp}) indeed gives the only terms allowed by symmetry, and in terms of $\mathcal{M}_{ij}$, they are:
\begin{equation}
    \begin{aligned}
    J_H &= \frac{1}{2}\left(\mathcal{M}_t^{11}+\mathcal{M}_t^{33}-\mathcal{M}_s^{11}-\mathcal{M}_s^{33}\right)\\
    J_{zz}^s &= \frac{1}{2}\left(\mathcal{M}_t^{33}-\mathcal{M}_t^{11}+\mathcal{M}_s^{11}-\mathcal{M}_s^{33}\right)\\ 
    J_{zz}^o &= \frac{1}{8}\left(\mathcal{M}_s^{33}-\mathcal{M}_s^{11}+3\mathcal{M}_t^{33}-3\mathcal{M}_t^{11}\right)\\ 
    J_{+} &= \frac{1}{2}\left({\mathcal{M}_t^{13}}^{\ast}-{\mathcal{M}_s^{13}}^{\ast}\right)\\
    J_{+-}^s &= \frac{1}{4}\left({\mathcal{M}_t^{12}}^{\ast}-{\mathcal{M}_s^{12}}^{\ast}\right)\\
    J_{++}^s &= \frac{1}{4}\left(\mathcal{M}_t^{34}-\mathcal{M}_s^{34}\right)\\
    J_{+-}^o &= \frac{1}{16}\left(3{\mathcal{M}_t^{12}}^{\ast}+{\mathcal{M}_s^{12}}^{\ast}\right)\\
    J_{++}^o &= \frac{1}{16}\left(3\mathcal{M}_t^{34}+\mathcal{M}_s^{34}\right)
    \end{aligned}
\end{equation}

We show how the couplings vary as a function of $a_M$ and $\varepsilon_r$ in Fig.~\ref{fig:J_aM}. The numerical values of the couplings for representative points in the phase diagram are shown in Table~\ref{tab:num_J}.

\begin{table}[htbp]
    \centering
    \footnotesize
    \setlength{\tabcolsep}{2pt} %
    \begin{tabularx}{\textwidth}{@{} l c *{8}{X} @{}}
        \toprule
        Phase & $(a_M, \varepsilon_r)$ & $J_H$ & $J_{zz}^s$ & $J_{zz}^o$ & $J_{+}$ & $J_{+-}^s$ & $J_{++}^s$ & $J_{+-}^o$ & $J_{++}^o$ \\
        \midrule
        FM & $(9.0, 5.0)$ & $-2.1$ & $-7.4$ & $-2.6$ & $3.9e^{-1.9i}$ & $3.8e^{0.9i}$ & $0.67$ & $0.4e^{0.9i}$ & $0.82$ \\
        \addlinespace
        Stripe & $(7.0, 5.0)$ & $-15.7$ & $-4.4$ & $-0.6$ & $21.0e^{-1.3i}$ & $16.5e^{1.4i}$ & $-6.0$ & $0.93e^{1.3i}$ & $0.1$ \\
        \addlinespace
        Zigzag & $(6.0, 8.0)$ & $0.3$ & $-17.7$ & $-2.8$ & $25.7e^{-1.7i}$ & $21.0e^{1.7i}$ & $8.8$ & $0.5e^{1.7i}$ & $2.7$ \\
        \addlinespace
        Nematic & $(10.0, 9.0)$ & $17.8$ & $-6.7$ & $-1.8$ & $16.1e^{2.4i}$ & $8.4e^{-1.4i}$ & $9.719$ & $1.7e^{-1.2i}$ & $2.850$ \\
        \addlinespace
        VBS & $(6.0, 10.0)$ & $7.3$ & $-29.3$ & $-5.7$ & $26.4e^{-1.8i}$ & $19.5e^{1.7i}$ & $15.1$ & $0.3e^{2.2i}$ & $4.4$ \\
        \addlinespace
        HSL & $(8.0, 7.0)$ & $2.9$ & $-16.2$ & $-4.2$ & $10.5e^{-2.0i}$ & $7.9e^{1.2i}$ & $4.5$ & $0.9e^{1.0i}$ & $2.1$ \\
        \bottomrule
    \end{tabularx}
    \caption{Numerical values of the couplings (in meV) at representative parameter points $(a_M, \varepsilon_r)$ for the different phases found in DMRG.}
    \label{tab:num_J}
\end{table}

\section{Defining electrical polarization operators via symmetry analysis}\label{app:pol}
In this section, we elaborate on the construction of the on-site and two-site (in-plane) electrical polarization operators. 
Quite generally, we can define the electric polarization $\bm P$ of a system as the coefficient in the first-order term $\bm P \cdot \bm E$ of expansion of the ground-state energy about zero field.  The free energy must be invariant under symmetries of the full Hamiltonian, so the polarization must transform in such a way as to make $\bm P \cdot \bm E$ a scalar under each of these.  The in-plane electric field $\bm{E}_\parallel$ transforms as a two-dimensional vector under $\mathcal{C}_{3v}$ and trivially under time-reversal $\mathcal{T}$ and spin-roations SU(2)$_s$, and the $z$-component $E^z$ transforms trivially under all symmetries.

In Table~I in the main text, we enumerated the action of the point-group and internal symmetries on the local operators $\eta^\pm$, $\eta^z$, and $\bm S$.  In Cartesian coordinates, these act on pseudospins as
\begin{align*}
    \mathcal{C}_{3z}&: \begin{pmatrix}\eta^x \\ \eta^y \\ \eta^z\end{pmatrix} \mapsto \begin{pmatrix}-1/2 & -\sqrt{3}/2 & 0 \\ \sqrt{3}/2 & -1/2 & 0 \\ 0 & 0 & 1 \end{pmatrix} 
    \begin{pmatrix}\eta^x \\ \eta^y \\ \eta^z\end{pmatrix} \\
    M_y&: (\eta^x,\eta^y,\eta^z) \mapsto (\eta^x,-\eta^y,-\eta^z)\\
    \mathcal{T}&: (\eta^x,\eta^y,\eta^z) \mapsto (\eta^x,\eta^y,-\eta^z).
\end{align*}
We note that the in-plane $\bm{\eta}_\parallel = (\eta^x, \eta^y)$ transforms as a vector under rotations, like $\bm{E}_\parallel$. 
However, unlike $\bm{E}_\parallel$ it behaves as a pseudo-vector under $M_y$; therefore, $\bm{\eta}_\parallel \cdot \bm{E}_\parallel$ is not a scalar operator under $\mathcal{C}_{3v}$. 
This can be fixed by defining $\bm P^{(1)}_i = \hat{z} \times \bm{\eta}_{\parallel,i} = \hat{z} \times \bm{\eta}_i$, which just rotates the in-plane component by ninety degrees.  
The polarization lacks an out-of-plane component as $\eta^z$ is odd under time-reversal. 
This is physically consistent with the charge density being restricted to the two-dimensional TMD plane, and out-of-planes displacements of the charge density are not considered in our effective model.

As mentioned in the main text, it is possible to go further if we allow two-site operators.  Indeed, the operator
\begin{equation*}
    \bm{P}^{(2)}_{ij} = (\eta^x_i \eta^y_j + \eta^y_i \eta^x_j , \eta^x_i \eta^x_j - \eta^y_i \eta^y_j)
\end{equation*}
transforms as a two-dimensional vector under $\mathcal{C}_{3v}$, just as $\bm{E}_\parallel$ does, so it makes a contribution to the polarization. For a rotation-invariant phase like HSL, where $\ev{\bm \eta} = 0$, this is the lowest-order symmetry-allowed contribution, as it can take different values on different bond directions, so long as the full configuration is invariant under the point group.  The magnitude of the coefficient, however, involves an overlap integral between Gaussian-localized electronic wavefunctions at different sites, so it is expected to be small.  In the main text, we refer to the two-site polarization $\bm{P}^{(2)}_{ij}$ as $\bm{P}_{ij}$ and write out $\hat z \times \bm \eta_\|$ explicitly for the one-site operator $\bm{P}^{(1)}$.

These operators can be interpreted in light of the fusion of irreps of $\mathcal C_{3v}$.  This group has three irreps: trivial $A_1$, pseudoscalar $A_2$, and vector $E$.  In order to consider two sites, we decompose the tensor product $E \otimes E = A_1 \oplus A_2 \oplus E$.  Of these, only the resulting $E$, which is $\bm{P}^{(2)}$, can couple to the electric field.

\section{Classical ground state optimization}\label{app:class}

\subsection{Setup and method}
In order to build an understanding of the model in Eq.(3) in the main text, we begin by looking the minimal energy ``ground state'' configurations of its classical limit, where the components of $\bm S$ and $\bm \eta$ are $c$-numbers.
We assume that the classical ground states retain some subgroup (the translation-breaking pattern) of the full plane group $p3m1$, and so are characterized by $\bm S$ and $\bm \eta$ on each of $n$ sublattices.
For a particular symmetry-breaking pattern, the classical Hamiltonian is a function $\mathcal{H}(\{\bm\eta_i\}, \{ \bm{S}_i\})$ from the product manifold $(S^2)^{\times 2n}$ of $2n$ two-spheres, one for pseudospin $\bm\eta$ and one for spin $\bm S$ on each sublattice, to the reals (the energy of the configuration).
Note that $\abs{\bm S} = 1/2$ and $\abs{\bm \eta} = 1.$
We then minimize this Hamiltonian function using manifold conjugate gradient descent \cite{kofod2025,edelman1998,absil2008}.
As a nonconvex optimization problem, this has no robust guarantees of convergence, but we run the optimization many times from random starting configurations in order to be confident that we've found the global minimum energy. 
This becomes more difficult for translation-breaking patterns with much more than twelve sites, but most of our optimal states are well below that size.
We can finally compare the lowest energies attained by each symmetry-breaking pattern to classify the phases by their symmetry, resulting in the classical phase diagram in Fig.~3(a) in the main text.  
In addition, we are able to compare the energies obtained from DMRG and classical optimization, as shown in Fig.~\ref{fig:vbs_opt}(a).

\subsection{Fully-classical optimization}
We probe translation-breaking patterns with up to twelve-site unit cells.  The crosshatched region of the classical phase diagram of Fig.~(3) in the main text is the region where six-, nine-, or twelve-site translation-breaking patterns have the lowest energy.  This suggests that either the true ground-state configuration has an even bigger unit cell or that its order is incommensurate with the lattice spacing.  Below we comment on the form of several of the classical phases in particular and discuss qualitatively how they arise from the couplings in Table~\ref{tab:num_J}.

\begin{figure}[t]
\centering
\includegraphics{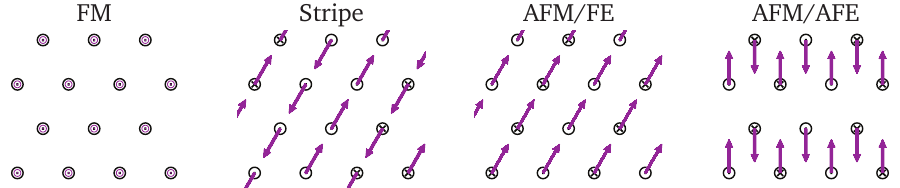}
\caption{The classical optimization finds four phases, two of which are also present in the quantum phase diagram. FM and Stripe are just the same as in Fig.~(3) in the main text. The new phases are antiferromagnetic/ferroelectric (AFM/FE) with a two-site unit cell and antiferromagnetic/antiferroelectric (AFM/AFE) with a four-site unit cell and rectangular translation symmetry.  Both of these phases include some degree of in-plane canting in some parts of the diagram, meaning that they can have noncollinear $\bm \eta$ (not shown in these representative configurations).  Both of them are overwhelmed in the quantum phase diagram by phases with entanglement, inaccessible to purely classical optimization.
}
\label{fig:class_configs}
\end{figure}

The Stripe and FM phases are the same as those that appear in the quantum model, and their configurations are shown in Fig.~3(a) in the main text.  There are two more phases that are only present in the classical configuration: an antiferromagnetic/ferroelectric (AFM/FE) and an antiferromagnetic/antiferroelectric (AFM/AFE)
.  These configurations are shown in Fig.~\ref{fig:class_configs}.  

\emph{Ferromagnet.}---The FM phase breaks time-reversal and exhibits a homogeneous spin alignment, preserving both the translation and point group symmetries. 
In our model, the orbital degrees of freedom also order, with $\eta^z \neq 0$ in both the quantum and classical settings.  We emphasize that such an orbital ordering does not signal a ferroelectric phase; unlike the in-plane component $\bm\eta_\parallel$ with is $\mathcal{T}$-symmetric,  $\eta^z$ maps to $-\eta^z$ under time-reversal.  
The dominant couplings at the FM points are $J_H$, $J_{zz}^s$, and $J_{zz}^o$, all of which are real and negative, preferring mutually reinforcing Heisenberg and Ising ferromagnets in the $\bm S$ and $\eta^z$ sectors, respectively.

\emph{Stripe.}---The origin of the stripe phase can be understood in terms of the dominant $H_+$ term, which couples linearly the in-plane orbital degrees of freedom with the spin-spin correlations. In this section, we argue that the Stripe configuration is a likely candidate to be the ground state of $H_+$ via a self-consistent optimization of the $\bm S$ and $\bm \eta$ degrees of freedom independently.

It is useful to write the Hamiltonian in terms of Cartesian spin coordinates $\bm\eta$ instead of ladder operators $\eta^\pm$ of Eq.~(3) in the main text.  
The couplings in the stripe region of the phase diagram is dominated by $J_+$, with $J_{+-}^s$ and $J_H$ also important.  These each multiply a Heisenberg term $\bm S_i \cdot \bm S_j$, and the $\bm \eta$ pieces of the $J_+$ and $J_{+-}^s$ contributions in Cartesian notation are
\begin{equation}\label{eq:cart_ham}
\begin{aligned}
    H_+ &= 2 \Re(J_+)\qty[\bm a_{ij}  \cdot (\bm\eta_i + \bm \eta_j)]
     + 2 \Im(J_+) \qty[ \bm a_{ij} \times \qty(\bm \eta_i - \bm \eta_j)] \cdot \hat z\\
    H_{+-}^s &= 2 \Re(J_{+-}^s ) \qty[\bm \eta_i \cdot \bm \eta_j - \eta^z_i \eta^z_j]
     + 2 \Im(J_{+-}^s) \qty[\bm \eta_i \times \bm\eta_j]\cdot\hat z
\end{aligned}
\end{equation}
where $\bm a_{ij}$ is the dimensionless bond vector (normalized by $a_M$) from site $i$ to site $j$. The imaginary parts of $J_+$ and $J_{+-}^s$ are much larger than their real parts in the Stripe region, so we will focus on these.

Firstly, we note that in order to take energetic advantage of the coupling with the largest magnitude, $J_+$, we can have neither uniform, ferroelectric $\bm \eta$ nor a uniform, ferromagnetic $\bm S$.  In the ferroelectric, the term $(\bm \eta_i - \bm \eta_j)$ would always be zero, while a ferromagnetic $\bm S$ configuration has uniform $\bm S_i \cdot \bm S_j$, meaning that the sum over $H_+$ is
\begin{equation}
    \sum_{\ev{ij}} H_+ \bm S_i \cdot \bm S_j \propto  \sum_{\ev{ij}} (\bm a_{ij} \times \bm \eta_i - \bm a_{ij} \times \bm \eta_j ) \cdot \hat z = 0,
\end{equation}
as each site appears an equal number of times as $i$ and as $j$.

Next we ask if there exists a configuration of spins and $\bm{\eta}$ which minimizes this term self-consistently. 
$H_+$ is linear in $\bm \eta$ and it couples to a Heisenberg term $\bm S_i \cdot \bm S_j$ that's quadratic in $\bm S$.  Under a fixed $\bm \eta$ configuration, this is a Heisenberg model for $\bm S$ with a bond-dependent coupling.  Under a fixed $\bm S$ configuration, $\bm \eta$ only experiences an ``external field'' generated by nonuniform $\bm S$ correlations. If $\bm S$ manages to be a ground state of the Heisenberg model generated by $\bm \eta$, and simultaneously $\bm \eta$ is parallel to the site-dependent field generated by $\bm S$, the state is self-consistent and stable under $H_+$.

In the observed ground state (Fig.~\ref{fig:class_configs}), we observe unit-period transverse ferroelectricity in $\bm \eta$ along with doubled-period antiferromagnetism in $\bm S$ with both ferromagnetic and antiferromagnetic correlations.  The Heisenberg coupling generated by $\bm \eta$ \emph{along} the stripe is zero.  Perpendicular to the stripes, the Heisenberg coupling is nonzero because $\bm \eta_i - \bm \eta_j$ isn't parallel to $\bm a_{ij}$, and it alternates sign along with $\bm \eta$.  This gives rise to the double-wide stripes in $\bm S$, with alternating spin-spin correlations---this is evidently the unique classical ground state of the effective Heisenberg model generated by the $\bm$ configuration up to $\mathrm{SU}(2)_s$.  

Any particular site $i$ has neighbors in the $\pm \bm a_{J}$ directions for $J=1,2,3$, which we'll denote here for brevity as $\bm S_{i \pm J}$. $H_+$ acts linearly on its orbital pseudospin $\bm \eta_i$, so we have
\begin{equation}
    2 \Im J_+ \sum_{\ev{ij}}  \hat z \cdot \qty(\bm a_{ij} \times (\bm \eta_i - \bm \eta_j))(\bm S_i \cdot \bm S_j)  = 2 \Im J_+ \sum_i \sum_{J=1}^{3}  (\bm S_i \cdot (\bm S_{i+J} - \bm S_{i-J})) (\hat z \times \bm a_J) \cdot \bm \eta_i.
\end{equation}
In our ground state configuration, the external field on $\bm \eta_i$ only receives contributions for bonds $J$ that are not in the stripe direction.  Each of these is in the direction $\hat z \times \bm a_J$, perpendicular to its bond, but the signs of the spin-spin correlations are such that in total, the components of the field perpendicular to the stripe cancel, and the $\bm \eta$ order is stabilized self-consistently.

We must also make sure that this order isn't strongly opposed or altered by $J_H$ or $J_{+-}^s$, the other two pertinent couplings.  The negative $J_H$ would prefer an $\bm S$-ferromagnet, but this is incompatible with the energetic gains to be made from $J_+$ as discussed above, so it must be satisfied that spin-spin correlations are ferromagnetic along two-thirds of the bonds.  The imaginary part of $J_{+-}^s$ multiplies the term $(\bm \eta_i \times \bm \eta_j)\cdot \hat z$, which is zero uniformly for the Stripe, as the $\bm \eta$ pattern is collinear.

\paragraph{AFM/FE and AFM/AFE.} We will not discuss these phases in detail, as they are less physically relevant.  
In particular, as discussed in the main text, we found that  these phases are destabilized by quantum fluctuations.
In the AFM region of the classical diagram, the $J_H$ term is dominant and antiferromagnetic, which leads to the overall frustrated antiferromagnetism in $\bm S$.  The role of the many other terms in setting the boundaries is less clear or important.

\subsection{Semiclassical optimization with a VBS ansatz.}\label{app:var_vbs}
\begin{figure}[t!]
\centering
\includegraphics[width=1.0\linewidth]{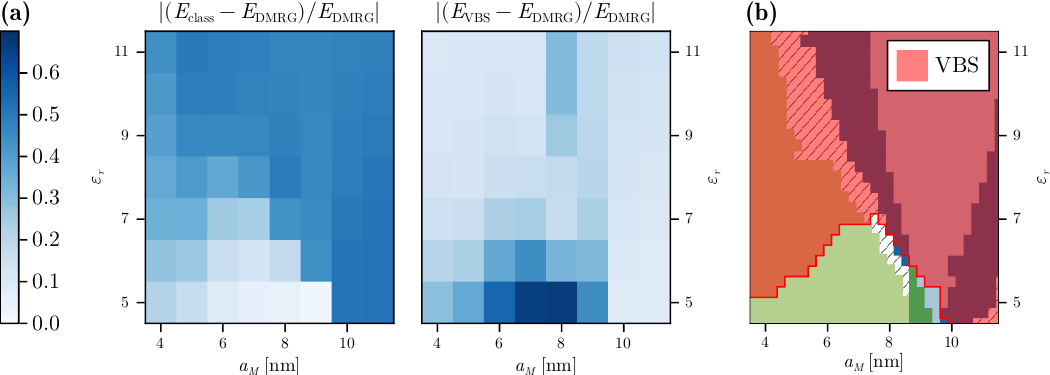}
\caption{(a) The classical ground states only give good approximations to DMRG energetics in the Stripe and FM phases.  
The VBS ansatz is able to get an energy similar to that from DMRG in the FiE-VBS, Nematic, and Zigzag regions. (b) The VBS semiclassical ansatz dominates the semiclassical phase diagram in all regions except near the Stripe and FM phases.  }
\label{fig:vbs_opt}
\end{figure}

Of the quantum phases, only Stripe and FM are directly accessible as classical configurations.  The Nematic and Zigzag phases are Luttinger-like, so they don't support a mean-field description. However, the FiE-VBS phase can still be examined semiclassically by fixing a pattern of $\bm S$-$\bm S$ singlets and then minimizing the energy over the $\bm \eta$ pseudospins: from the perspective of classical $\bm \eta$, it is no object that the singlet correlations $\ev{\bm S_i \cdot \bm S_j} = -3/4$ are able to exceed classical bounds, and the optimization can proceed in the same way. When we do this, we see that the VBS configuration has a lower energy than any other classical configuration in most of the phase diagram---all of it except the Stripe and FM regions, as shown in Fig.~\ref{fig:vbs_opt}.  The $\langle \bm \eta \rangle$ pattern is the same as what we see in iDMRG.  Note that the Helical SL phase emerges in the region of the semi-classical phase diagram that borders Stripe, FM, FiE-VBS, and the unconverged sliver, in line with our expectation that such a state be enabled by great frustration.

We also use this semiclassical approach to test ``brick wall'' VBS patterns, where the singlets are all aligned in the same direction.  Neither of the two distinct brick-wall VBS ans\"atze have a favorable energy anywhere in the diagram.  This supports the conclusion that the Nematic and Zigzag states that we see with DMRG are truly Luttinger-like, instead of just being a cat state of translation-breaking VBS states along the short axis of the the cylinder.

\section{DMRG methodology and results}
\subsection{Methods}
To find the quantum phase diagram in Fig.~(3), we use the infinite density-matrix renormalization group (iDMRG) as implemented in TenPy \cite{tenpy2024}, performed on a cylinder of circumference $L_y = 4,6$ and with $\mathrm{U}(1)_s$-symmetric tensors of bond dimension up to $\chi = 2560$.  

\paragraph{Cylinder geometry.} We wrap the triangular lattice such that the $\pm \bm a_1$ direction runs along the cylinder and the $\pm \bm a_3$ direction wraps around it: a translation of $\pm L_y \bm a_3$ carries a site back to where it started. We use a matrix product state unit cell of of $L_x L_y$ independent tensors, where $(L_x,L_y) = (2,4)$ or $(1,6)$.
This cylinder geometry breaks $\mathcal{C}_{3v}$, of course, but it actually preserves the mirror that maps $\bm a_3 \mapsto - \bm a_3$, which we will notate as $M_3$.  We run DMRG on all points for $L_y=4$, but we make additional runs at $L_y=6$ for points near phase boundaries and to make sure that our phases are not being ``trapped'' by the finite cylinder size, which lets us get a better handle on the physical picture in the thermodynamic limit.

\paragraph{Pinning fields.} The Mermin-Wagner theorem forbids spontaneous breaking of a continuous symmetry in a classical two-dimensional statistical model at finite temperature \cite{AuerbachBook}.  By the quantum-to-classical mapping, it can be shown that the quantum path integral that generates the ground-state properties of our model on a finite-circumference infinite-length cylinder is equivalent to a classical thermal path integral for a two-dimensional system at finite temperature---under a Wick rotation, the finite circumference is goes to a Euclidean time/inverse temperature $\beta$, and the infinite time dimension becomes an infinite spatial dimension.  Therefore, we should not expect to observe states that break $\mathrm{SU}(2)_s$ in DMRG, even if they would actually be symmetry-broken in the thermodynamic limit of a fully two-dimensional quantum system with infinite extent.

This is relevant to our work only in the Stripe and Helical SL phases, both of which break $\mathrm{SU}(2)_s$. While FM breaks $\mathrm{SU}(2)_s$, we are able to see this directly by using tensors that fix the overall $S^z$ sector to be fully-polarized.  The difference is that while Stripe and Helical SL break $\mathrm{SU}(2)_s$, their order parameters do not themselves generate a symmetry of the Hamiltonian, so they do not generate suitable conserved quantities for quantum number-conserving DMRG. 
One possible solution is to use a pinning field (e.g., as done in Ref.~\cite{SD2024} to see AFM order): we add a term to the Hamiltonian which couples directly to the order parameter that we are interested in, which allows us to measure its susceptibility. In a disordered state, we would expect the susceptibility to be finite and analytic as the field is swept through zero. If the susceptibility diverges near zero applied field, we have a state with long-range order, even if DMRG is not able to resolve that order in a quasi-one-dimensional setting.

\begin{figure}
    \centering
    \includegraphics[width=0.8\linewidth]{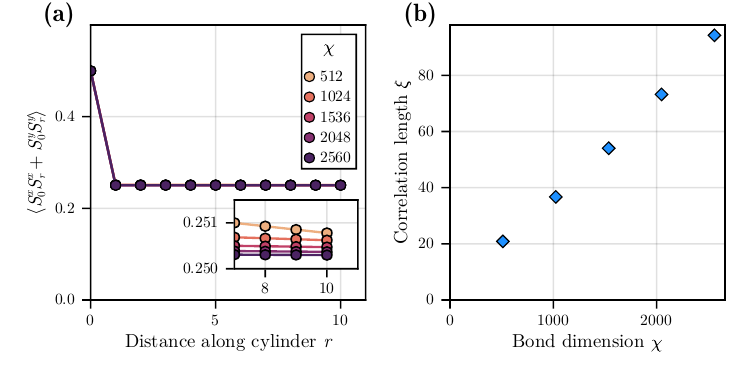}
    \caption{
    (a) Two-point spin-spin correlation function for $(a_M, \epsilon_r) = (9.0~\mathrm{nm}, 5.0)$ for a $L_y=4$ cylinder. We observe that the system exhibits correlations that decay extremely slowly with distance $r$. Further, the decay of correlations is suppressed as the bond dimension $\chi$ increases (see inset).
    (b) Extracted spin-spin correlation length $\xi$ measured along the cylinder, as a function of bond dimension $\chi$. We explicitly observe an increasing correlation length with increasing bond dimension, strongly indicating that spin-rotation symmetry is broken in the thermodynamic limit.
    }
    \label{fig:app_FM}
\end{figure}

\subsection{Results by phase}
\subsubsection{Ferromagnet (FM)} The FM phase breaks time-reversal $\mathcal{T}$ and spin-rotation $\mathrm{SU}(2)_s$ while preserving the space group. 
In the sector with total spin $S^z = 0$, the iDMRG wavefunction is unable to express a nonzero local expectation value $\ev{\bm S_\|}$ since $\bm{S}_\|$ is an operator without definite U(1) charge.
Nevertheless, the FM order will be manifest in the present of long-range spin-spin correlations, $\langle S^x_0S^x_r +  S^y_0S^y_r\rangle$~[Fig.~\ref{fig:app_FM}(a)].
However, one must be careful---the iDMRG formally defines an effective one dimensional quantum problem that cannot, via the Mermim-Wagner theorem, exhibit long-range order when breaking a continuous symmetry.
Indeed, we observe that the correlation length of the system remains finite and grows with increasing bond dimension~[Fig.~\ref{fig:app_FM}(b)].
These are known signatures of continuous symmetry breaking in iDMRG.
With respect to the orbital degrees of freedom, we observe almost complete polarization along the orbital pseudospin quantization axis, with $\ev{\eta^z} = 0.97294$.

\subsubsection{Stripe} 
\begin{figure}
[t!]
\centering
\includegraphics{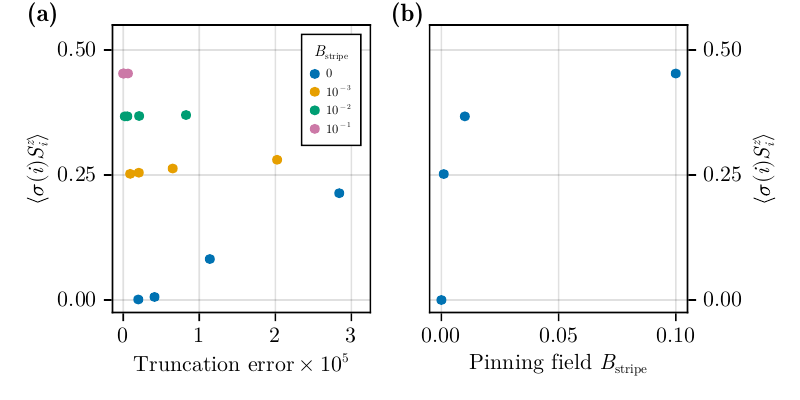}
\caption{Stripe order parameter as a function of pinning field at parameters $(a_M,\epsilon_r) = (7.0~\mathrm{nm},5.0)$, with $\sigma(i) = \cos (\bm r_i \cdot \bm Q_s) + \sin (\bm r_i \cdot \bm Q_s)$ as in Eq.~\ref{eq:stripe_pin}.  Under a pinning field $B_{\text{stripe}}$, the susceptibility diverges, indicating spontaneous symmetry breaking, even for a continuous symmetry of a quasi-one-dimensional system.  The values in (b) are linear extrapolations to zero truncation error of the measured data in (a).} 
\label{fig:stripe_pin}
\end{figure}

The Stripe phase has a four-site unit cell, breaking the point group $\mathcal{C}_{3v}$, time-reversal $\mathcal{T}$, and spin rotation $\mathrm{SU}(2)_s$.  Our Stripe ordering is around the cylinder in the $\hat a_3$ direction. The antiferroelectric order parameter is almost classical at $\ev{\hat{a}_3 \cdot \bm \eta} = \pm 0.983$.  Because the Stripe phase breaks the continuous symmetry $\mathrm{SU}(2)_s$, we observe the symmetry breaking by applying a straightforward pinning field of the form
\begin{equation}\label{eq:stripe_pin}
H_{\text{stripe}} =B_{\text{stripe}}\sum_i \sigma(i) S_i^z =  B_{\text{stripe}}\sum_i \qty(\cos(\bm r_i \cdot \bm Q_s) + \sin(\bm r_i \cdot \bm Q_s))S_i^z ,
\end{equation}
where $\bm r_i$ is the position of the $i$th lattice point and $\bm Q_s = \pi(\hat{z} \times \bm{a}_3)/(\sqrt{3} a_M^2) = \frac{\pi}{2a_M}\hat x - \frac{\pi}{2\sqrt{3}a_M} \hat y$.  This favors the Stripe order on $\bm S$.  We extrapolate on truncation error to estimate the Stripe order parameter at infinite bond dimension.  These results are presented in  Fig.~\ref{fig:stripe_pin}, which demonstrates that the susceptibility $\chi_{\text{stripe}} = \pdv{ \ev{\sigma(i) S_i^z}}{B_{\text{stripe}}}$ is very large at zero field.  This data is computed for $L_y = 4$, but we expect the curve to become a step function in the thermodynamic limit as $L_y$ goes to infinity.

\subsubsection{Ferrielectric Valence Bond Solid (FiE-VBS)}
\begin{figure}
[t!]
\centering
\includegraphics{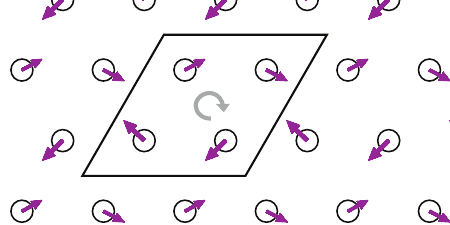}
\caption{After subtracting off the net ferroelectric moment, the remaining $\ev{\bm \eta_\parallel}$ of the FiE-VBS state approaches a four-sublattice $90^\circ$ ordering pattern, with the unit cell and direction of chirality shown.} 
\label{fig:supp_vbs_afe}
\end{figure}

The FiE-VBS state breaks only space group symmetries, maintaining time-reversal $\mathcal{T}$ and spin-rotation $\mathrm{SU}(2)_s$.  We measure the spin correlations between bonded sites as $\ev{\bm S_i \cdot \bm S_j} = -0.568$, reduced in magnitude from the $-3/4$ of an ideal singlet.  The state is ferrielectric because it has a net average moment per site of $\ev{\bm \eta}_i = 0.152 \hat a_1$.  The residual antiferroelectricity after this moment is subtracted off is close to a four-sublattice $90^\circ$ pattern [Fig.~\ref{fig:supp_vbs_afe}].

\subsubsection{Helical Spin Liquid (HSL)}\label{app:hsl}
The Helical SL phase also breaks $\mathrm{SU}(2)_s$, but we do not see the susceptibility diverge as cleanly as we do in the Stripe.  The HSL pinning field is
\begin{equation}
    H_{\text{HSL}} = B_{\text{HSL}} \sum_{\ev{ij}} C^z_{ij} = B_{\text{HSL}} \sum_{\ev{ij}} (\bm{S}_i \times \bm{S}_j)^z,
\end{equation}
which is notably different form $H_{\text{stripe}}$ in that the terms on different bonds in the external field Hamiltonian don't commute with each other. This means that we don't have a clear picture of a mean-field Hamiltonian with a simple, unentangled ground state, as we do in most cases of spontaneous symmetry breaking.  That being said, we can still check the dependence of the order parameter on the field to check for signs of symmetry-breaking.

\begin{figure}
[t!]
\centering
\includegraphics{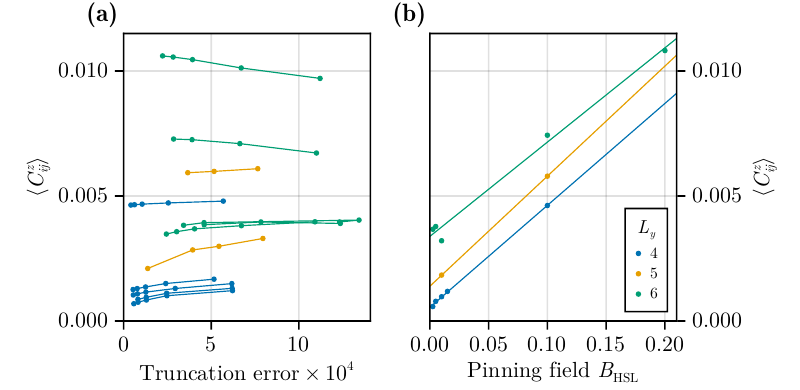}
\caption{Similarly to the Stripe case, the HSL order at $(a_M, \epsilon_r) = (8.0~\mathrm{nm}, 7.0)$ can be stabilized by adding a pinning field $B_{HSL}$.  In (a), the measured order parameter $\ev{C^z_{ij}}$ is plotted as a function of truncation error for various pinning fields at $L_y = 4,5, \text{ and } 6$, and the extrapolant at zero truncation error is plotted as a function of bond dimension in (b).  The approach to zero is not as linear as in the Stripe, but it is still clear that the extrapolated order parameter increases with increasing $L_y$.} 
\label{fig:hsl_pin}
\end{figure}

The results are shown in Fig.~\ref{fig:hsl_pin}.  
At moderate fields, the helical order parameter converges cleanly in truncation error to a finite extrapolant, as is the case for the Stripe.  
At smaller fields, the approach is more nonlinear, so we cannot be as confident about our extrapolation.  For the fields down to $B_{HSL} = 0.0025$ that we have investigated, though, the order parameter seems to converge, and above all else it is clear that the value of the order parameter is significantly higher for $L_y = 6$ than $L_y = 4$.  This supports our claim that the operator $C_{ij}^z$ orders spontaneously in the full $L_y \to \infty$ thermodynamic limit.

The cylinder geometry explicitly breaks the point group symmetry $\mathcal{C}_{3v}$, but it preserves the $\bm a_3 \mapsto -\bm a_3$ mirror $M_3$, as discussed above.  Due to this finite-size effect, that our DMRG results display a small in-plane pseudospin expectation value $\ev{\bm \eta_\parallel}$ which is approximately proportional to $\bm a_3$. The $M_3$-breaking component perpendicular to $\bm a_3$ goes to zero with truncation error, and the magnitude of this vector goes to zero as the cylinder circumference goes to infinity and full $\mathcal{C}_{3v}$ is restored.  For $L_y=6$ at $\chi = 2560$ in particular, the average $M_3$-preserving component $\ev{\bm \eta_\parallel} \cdot \hat a_3$ is $2.1 \times 10^{-2}$, and the average $M_3$-breaking component $|\ev{\bm \eta_\parallel} \times \hat a_3)|$ is $1.9 \times 10^{-4}$, where $\hat a_3 = \bm a_3 / a_M$.

\end{appendix}

\bibliography{Refs}

@article{wuHubbardModelPhysics2018,
  title = {Hubbard {{Model Physics}} in {{Transition Metal Dichalcogenide Moir}}\'e {{Bands}}},
  author = {Wu, Fengcheng and Lovorn, Timothy and Tutuc, Emanuel and MacDonald, A. H.},
  year = {2018},
  month = jul,
  journal = {Phys. Rev. Lett.},
  volume = {121},
  number = {2},
  pages = {026402},
  publisher = {{American Physical Society}},
  doi = {10.1103/PhysRevLett.121.026402}
}

@article{Reddy2023,
  title = {Artificial Atoms, Wigner Molecules, and an Emergent Kagome Lattice in Semiconductor Moir\'e Superlattices},
  author = {Reddy, Aidan P. and Devakul, Trithep and Fu, Liang},
  journal = {Phys. Rev. Lett.},
  volume = {131},
  issue = {24},
  pages = {246501},
  numpages = {6},
  year = {2023},
  month = {Dec},
  publisher = {American Physical Society},
  doi = {10.1103/PhysRevLett.131.246501},
  url = {https://link.aps.org/doi/10.1103/PhysRevLett.131.246501}
}

@ARTICLE{Luo2023,
       author = {{Luo}, Di and {Reddy}, Aidan P. and {Devakul}, Trithep and {Fu}, Liang},
        title = "{Artificial intelligence for artificial materials: moir{\'e} atom}",
      journal = {arXiv e-prints},
         year = 2023,
        month = mar,
          eid = {arXiv:2303.08162},
          doi = {10.48550/arXiv.2303.08162},
archivePrefix = {arXiv},
       eprint = {2303.08162},
 primaryClass = {cond-mat.str-el},
       adsurl = {https://ui.adsabs.harvard.edu/abs/2023arXiv230308162L}
}

@article{Mikhailov:2002,
  title = {Quantum-dot lithium in zero magnetic field: Electronic properties, thermodynamics, and Fermi liquid--Wigner solid crossover in the ground state},
  author = {Mikhailov, S. A.},
  journal = {Phys. Rev. B},
  volume = {65},
  issue = {11},
  pages = {115312},
  numpages = {12},
  year = {2002},
  month = {Feb},
  publisher = {American Physical Society},
  doi = {10.1103/PhysRevB.65.115312},
  url = {https://link.aps.org/doi/10.1103/PhysRevB.65.115312}
}

@article{PhysRevB.108.L121411,
  title = {Quantum Wigner molecules in moir\'e materials},
  author = {Yannouleas, Constantine and Landman, Uzi},
  journal = {Phys. Rev. B},
  volume = {108},
  issue = {12},
  pages = {L121411},
  numpages = {7},
  year = {2023},
  month = {Sep},
  publisher = {American Physical Society},
  doi = {10.1103/PhysRevB.108.L121411},
  url = {https://link.aps.org/doi/10.1103/PhysRevB.108.L121411}
}

@article{lopez2000matrix,
  title={Matrix elements for the one-dimensional harmonic oscillator},
  author={L{\'o}pez-Bonilla, J and Ovando, G},
  journal={Bull. Irish Math. Soc.(44)},
  volume={61},
  doi = {https://www.irishmathsoc.org/bull44/oscillator.pdf},
  year={2000}
}

@article{GUSEINOV2006226,
title = {Exact analytical expressions and numerical analysis of two-center Franck–Condon factors and matrix elements over displaced harmonic oscillator wave functions},
journal = {Computer Physics Communications},
volume = {175},
number = {3},
pages = {226-231},
year = {2006},
issn = {0010-4655},
doi = {https://doi.org/10.1016/j.cpc.2006.04.002},
url = {https://www.sciencedirect.com/science/article/pii/S0010465506001603},
author = {I.I. Guseinov and B.A. Mamedov and A.S. Ekenoğlu},
}

@article{PhysRevResearch.2.033087,
  title = {Band topology, Hubbard model, Heisenberg model, and Dzyaloshinskii-Moriya interaction in twisted bilayer ${\mathrm{WSe}}_{2}$},
  author = {Pan, Haining and Wu, Fengcheng and Das Sarma, Sankar},
  journal = {Phys. Rev. Res.},
  volume = {2},
  issue = {3},
  pages = {033087},
  numpages = {13},
  year = {2020},
  month = {Jul},
  publisher = {American Physical Society},
  doi = {10.1103/PhysRevResearch.2.033087},
  url = {https://link.aps.org/doi/10.1103/PhysRevResearch.2.033087}
}

@article{PhysRevB.102.201104,
  title = {Quantum phase diagram of a Moir\'e-Hubbard model},
  author = {Pan, Haining and Wu, Fengcheng and Das Sarma, Sankar},
  journal = {Phys. Rev. B},
  volume = {102},
  issue = {20},
  pages = {201104},
  numpages = {5},
  year = {2020},
  month = {Nov},
  publisher = {American Physical Society},
  doi = {10.1103/PhysRevB.102.201104},
  url = {https://link.aps.org/doi/10.1103/PhysRevB.102.201104}
}

@article{PhysRevB.102.235423,
  title = {Charge transfer excitations, pair density waves, and superconductivity in moir\'e materials},
  author = {Slagle, Kevin and Fu, Liang},
  journal = {Phys. Rev. B},
  volume = {102},
  issue = {23},
  pages = {235423},
  numpages = {11},
  year = {2020},
  month = {Dec},
  publisher = {American Physical Society},
  doi = {10.1103/PhysRevB.102.235423},
  url = {https://link.aps.org/doi/10.1103/PhysRevB.102.235423}
}

@article{PhysRevLett.122.086402,
  title = {Topological Insulators in Twisted Transition Metal Dichalcogenide Homobilayers},
  author = {Wu, Fengcheng and Lovorn, Timothy and Tutuc, Emanuel and Martin, Ivar and MacDonald, A. H.},
  journal = {Phys. Rev. Lett.},
  volume = {122},
  issue = {8},
  pages = {086402},
  numpages = {5},
  year = {2019},
  month = {Feb},
  publisher = {American Physical Society},
  doi = {10.1103/PhysRevLett.122.086402},
  url = {https://link.aps.org/doi/10.1103/PhysRevLett.122.086402}
}

@Article{LiQAH2021,
author={Li, Tingxin
and Jiang, Shengwei
and Shen, Bowen
and Zhang, Yang
and Li, Lizhong
and Tao, Zui
and Devakul, Trithep
and Watanabe, Kenji
and Taniguchi, Takashi
and Fu, Liang
and Shan, Jie
and Mak, Kin Fai},
title={Quantum anomalous Hall effect from intertwined moir{\'e} bands},
journal={Nature},
year={2021},
month={Dec},
day={01},
volume={600},
number={7890},
pages={641-646},
issn={1476-4687},
doi={10.1038/s41586-021-04171-1},
url={https://doi.org/10.1038/s41586-021-04171-1}
}

@Article{Cai2023,
author={Cai, Jiaqi
and Anderson, Eric
and Wang, Chong
and Zhang, Xiaowei
and Liu, Xiaoyu
and Holtzmann, William
and Zhang, Yinong
and Fan, Fengren
and Taniguchi, Takashi
and Watanabe, Kenji
and Ran, Ying
and Cao, Ting
and Fu, Liang
and Xiao, Di
and Yao, Wang
and Xu, Xiaodong},
title={Signatures of fractional quantum anomalous Hall states in twisted MoTe2},
journal={Nature},
year={2023},
month={Oct},
day={01},
volume={622},
number={7981},
pages={63-68},
issn={1476-4687},
doi={10.1038/s41586-023-06289-w},
url={https://doi.org/10.1038/s41586-023-06289-w}
}

@article{PhysRevX.13.031037,
  title = {Observation of Integer and Fractional Quantum Anomalous Hall Effects in Twisted Bilayer ${\mathrm{MoTe}}_{2}$},
  author = {Xu, Fan and Sun, Zheng and Jia, Tongtong and Liu, Chang and Xu, Cheng and Li, Chushan and Gu, Yu and Watanabe, Kenji and Taniguchi, Takashi and Tong, Bingbing and Jia, Jinfeng and Shi, Zhiwen and Jiang, Shengwei and Zhang, Yang and Liu, Xiaoxue and Li, Tingxin},
  journal = {Phys. Rev. X},
  volume = {13},
  issue = {3},
  pages = {031037},
  numpages = {12},
  year = {2023},
  month = {Sep},
  publisher = {American Physical Society},
  doi = {10.1103/PhysRevX.13.031037},
  url = {https://link.aps.org/doi/10.1103/PhysRevX.13.031037}
}

@article{PhysRevX.14.011004,
  title = {Valley-Coherent Quantum Anomalous Hall State in AB-Stacked ${\mathrm{MoTe}}_{2}/{\mathrm{W}\mathrm{S}\mathrm{e}}_{2}$ Bilayers},
  author = {Tao, Zui and Shen, Bowen and Jiang, Shengwei and Li, Tingxin and Li, Lizhong and Ma, Liguo and Zhao, Wenjin and Hu, Jenny and Pistunova, Kateryna and Watanabe, Kenji and Taniguchi, Takashi and Heinz, Tony F. and Mak, Kin Fai and Shan, Jie},
  journal = {Phys. Rev. X},
  volume = {14},
  issue = {1},
  pages = {011004},
  numpages = {8},
  year = {2024},
  month = {Jan},
  publisher = {American Physical Society},
  doi = {10.1103/PhysRevX.14.011004},
  url = {https://link.aps.org/doi/10.1103/PhysRevX.14.011004}
}

@article{PhysRevLett.82.5325,
  title = {Spontaneous Symmetry Breaking in Single and Molecular Quantum Dots},
  author = {Yannouleas, Constantine and Landman, Uzi},
  journal = {Phys. Rev. Lett.},
  volume = {82},
  issue = {26},
  pages = {5325--5328},
  numpages = {0},
  year = {1999},
  month = {Jun},
  publisher = {American Physical Society},
  doi = {10.1103/PhysRevLett.82.5325},
  url = {https://link.aps.org/doi/10.1103/PhysRevLett.82.5325}
}

@article{PhysRevLett.82.3320,
  title = {Crossover from Fermi Liquid to Wigner Molecule Behavior in Quantum Dots},
  author = {Egger, R. and H\"ausler, W. and Mak, C. H. and Grabert, H.},
  journal = {Phys. Rev. Lett.},
  volume = {82},
  issue = {16},
  pages = {3320--3323},
  numpages = {0},
  year = {1999},
  month = {Apr},
  publisher = {American Physical Society},
  doi = {10.1103/PhysRevLett.82.3320},
  url = {https://link.aps.org/doi/10.1103/PhysRevLett.82.3320}
}

@article{PhysRevLett.96.126806,
  title = {Excitation Spectrum of Two Correlated Electrons in a Lateral Quantum Dot with Negligible Zeeman Splitting},
  author = {Ellenberger, C. and Ihn, T. and Yannouleas, C. and Landman, U. and Ensslin, K. and Driscoll, D. and Gossard, A. C.},
  journal = {Phys. Rev. Lett.},
  volume = {96},
  issue = {12},
  pages = {126806},
  numpages = {4},
  year = {2006},
  month = {Mar},
  publisher = {American Physical Society},
  doi = {10.1103/PhysRevLett.96.126806},
  url = {https://link.aps.org/doi/10.1103/PhysRevLett.96.126806}
}

@Article{Kalliakos2008,
author={Kalliakos, Sokratis
and Rontani, Massimo
and Pellegrini, Vittorio
and Garc{\'i}a, C{\'e}sar Pascual
and Pinczuk, Aron
and Goldoni, Guido
and Molinari, Elisa
and Pfeiffer, Loren N.
and West, Ken W.},
title={A molecular state of correlated electrons in a quantum dot},
journal={Nature Physics},
year={2008},
month={Jun},
day={01},
volume={4},
number={6},
pages={467-471},
issn={1745-2481},
doi={10.1038/nphys944},
url={https://doi.org/10.1038/nphys944}
}

@article{
doi:10.1126/science.adk1348,
author = {Hongyuan Li  and Ziyu Xiang  and Aidan P. Reddy  and Trithep Devakul  and Renee Sailus  and Rounak Banerjee  and Takashi Taniguchi  and Kenji Watanabe  and Sefaattin Tongay  and Alex Zettl  and Liang Fu  and Michael F. Crommie  and Feng Wang },
title = {Wigner molecular crystals from multielectron moiré artificial atoms},
journal = {Science},
volume = {385},
number = {6704},
pages = {86-91},
year = {2024},
doi = {10.1126/science.adk1348},
URL = {https://www.science.org/doi/abs/10.1126/science.adk1348}
}

@article{
Kugel_1982,
doi = {10.1070/PU1982v025n04ABEH004537},
url = {https://dx.doi.org/10.1070/PU1982v025n04ABEH004537},
year = {1982},
month = {apr},
publisher = {},
volume = {25},
number = {4},
pages = {231},
author = {Kliment I Kugel' and D I Khomskiĭ},
title = {The Jahn-Teller effect and magnetism: transition metal
compounds},
journal = {Soviet Physics Uspekhi}
}

@article{PhysRevB.109.L121302,
  title = {Wigner-molecule supercrystal in transition metal dichalcogenide moir\'e superlattices: Lessons from the bottom-up approach},
  author = {Yannouleas, Constantine and Landman, Uzi},
  journal = {Phys. Rev. B},
  volume = {109},
  issue = {12},
  pages = {L121302},
  numpages = {7},
  year = {2024},
  month = {Mar},
  publisher = {American Physical Society},
  doi = {10.1103/PhysRevB.109.L121302},
  url = {https://link.aps.org/doi/10.1103/PhysRevB.109.L121302}
}

@article{PhysRevLett.133.246502,
  title = {Crystal-Field Effects in the Formation of Wigner-Molecule Supercrystals in Moir\'e Transition Metal Dichalcogenide Superlattices},
  author = {Yannouleas, Constantine and Landman, Uzi},
  journal = {Phys. Rev. Lett.},
  volume = {133},
  issue = {24},
  pages = {246502},
  numpages = {9},
  year = {2024},
  month = {Dec},
  publisher = {American Physical Society},
  doi = {10.1103/PhysRevLett.133.246502},
  url = {https://link.aps.org/doi/10.1103/PhysRevLett.133.246502}
}

@article{li2024emergent,
  title={Emergent Wigner phases in moir$\backslash$'e superlattice from deep learning},
  author={Li, Xiang and Qian, Yubing and Ren, Weiluo and Xu, Yang and Chen, Ji},
  journal={arXiv preprint arXiv:2406.11134},
  doi={https://doi.org/10.48550/arXiv.2406.11134},
  url={https://doi.org/10.48550/arXiv.2406.11134},
  year={2024}
}

@article{Li_2025,
doi = {10.1088/1674-1056/ad9ffc},
url = {https://dx.doi.org/10.1088/1674-1056/ad9ffc},
year = {2025},
month = {jan},
publisher = {Chinese Physical Society and IOP Publishing Ltd},
volume = {34},
number = {2},
pages = {027303},
author = {Li, Yanqi and Wang, Yi-Jie and Song, Zhi-Da},
title = {Orbital XY models in moiré superlattices},
journal = {Chinese Physics B}
}

@article{PhysRevB.109.045144,
  title = {Itinerant ferromagnetism in transition metal dichalcogenide moir\'e superlattices},
  author = {Potasz, Pawel and Morales-Dur\'an, Nicol\'as and Hu, Nai Chao and MacDonald, Allan H.},
  journal = {Phys. Rev. B},
  volume = {109},
  issue = {4},
  pages = {045144},
  numpages = {10},
  year = {2024},
  month = {Jan},
  publisher = {American Physical Society},
  doi = {10.1103/PhysRevB.109.045144},
  url = {https://link.aps.org/doi/10.1103/PhysRevB.109.045144}
}

@article{PhysRevB.107.235131,
  title = {Magnetism and quantum melting in moir\'e-material Wigner crystals},
  author = {Morales-Dur\'an, Nicol\'as and Potasz, Pawel and MacDonald, Allan H.},
  journal = {Phys. Rev. B},
  volume = {107},
  issue = {23},
  pages = {235131},
  numpages = {11},
  year = {2023},
  month = {Jun},
  publisher = {American Physical Society},
  doi = {10.1103/PhysRevB.107.235131},
  url = {https://link.aps.org/doi/10.1103/PhysRevB.107.235131}
}

@article{
doi:10.1073/pnas.2021826118,
author = {Mattia Angeli  and Allan H. MacDonald },
title = {$\mathrm{\ensuremath{\Gamma}}$ valley transition metal dichalcogenide moiré bands},
journal = {Proceedings of the National Academy of Sciences},
volume = {118},
number = {10},
pages = {e2021826118},
year = {2021},
doi = {10.1073/pnas.2021826118},
URL = {https://www.pnas.org/doi/abs/10.1073/pnas.2021826118},
eprint = {https://www.pnas.org/doi/pdf/10.1073/pnas.2021826118}
}

@article{PhysRevLett.120.107703,
  title = {Tunable $\mathrm{\ensuremath{\Gamma}}\ensuremath{-}K$ Valley Populations in Hole-Doped Trilayer ${\mathrm{WSe}}_{2}$},
  author = {Movva, Hema C. P. and Lovorn, Timothy and Fallahazad, Babak and Larentis, Stefano and Kim, Kyounghwan and Taniguchi, Takashi and Watanabe, Kenji and Banerjee, Sanjay K. and MacDonald, Allan H. and Tutuc, Emanuel},
  journal = {Phys. Rev. Lett.},
  volume = {120},
  issue = {10},
  pages = {107703},
  numpages = {5},
  year = {2018},
  month = {Mar},
  publisher = {American Physical Society},
  doi = {10.1103/PhysRevLett.120.107703},
  url = {https://link.aps.org/doi/10.1103/PhysRevLett.120.107703}
}

@article{brzezinska2024pressure,
doi = {10.1088/2053-1583/ad7c5f},
url = {https://doi.org/10.1088/2053-1583/ad7c5f},
year = {2024},
month = {oct},
publisher = {IOP Publishing},
volume = {12},
number = {1},
pages = {015003},
author = {Brzezińska, Marta and Grytsiuk, Sergii and Rösner, Malte and Gibertini, Marco and Rademaker, Louk},
title = {Pressure-tuned many-body phases through $\mathrm{\ensuremath{\Gamma}}$-K valleytronics in moiré bilayer WSe2},
journal = {2D Materials},
}

@article{PhysRevLett.131.046401,
  title = {Flat $\mathrm{\ensuremath{\Gamma}}$ Moir\'e Bands in Twisted Bilayer ${\mathrm{WSe}}_{2}$},
  author = {Gatti, G. and Issing, J. and Rademaker, L. and Margot, F. and de Jong, T. A. and van der Molen, S. J. and Teyssier, J. and Kim, T. K. and Watson, M. D. and Cacho, C. and Dudin, P. and Avila, J. and Edwards, K. Cordero and Paruch, P. and Ubrig, N. and Guti\'errez-Lezama, I. and Morpurgo, A. F. and Tamai, A. and Baumberger, F.},
  journal = {Phys. Rev. Lett.},
  volume = {131},
  issue = {4},
  pages = {046401},
  numpages = {7},
  year = {2023},
  month = {Jul},
  publisher = {American Physical Society},
  doi = {10.1103/PhysRevLett.131.046401},
  url = {https://link.aps.org/doi/10.1103/PhysRevLett.131.046401}
}

@article{Zheng2006,
  title = {Anomalous Excitation Spectra of Frustrated Quantum Antiferromagnets},
  author = {Zheng, Weihong and Fj\ae{}restad, John O. and Singh, Rajiv R. P. and McKenzie, Ross H. and Coldea, Radu},
  journal = {Phys. Rev. Lett.},
  volume = {96},
  issue = {5},
  pages = {057201},
  numpages = {4},
  year = {2006},
  month = {Feb},
  publisher = {American Physical Society},
  doi = {10.1103/PhysRevLett.96.057201},
  url = {https://link.aps.org/doi/10.1103/PhysRevLett.96.057201}
}

@article{Hayashi2007,
author = {Hayashi ,Yuta and Ogata ,Masao},
title = {Possibility of Gapless Spin Liquid State by One-Dimensionalization},
journal = {Journal of the Physical Society of Japan},
volume = {76},
number = {5},
pages = {053705},
year = {2007},
doi = {10.1143/JPSJ.76.053705},

URL = { 
    
        https://doi.org/10.1143/JPSJ.76.053705
    
    

},
eprint = { 
    
        https://doi.org/10.1143/JPSJ.76.053705
    
    

}
}

@article{Heidarian2009,
  title = {Spin-$\frac{1}{2}$ Heisenberg model on the anisotropic triangular lattice: From magnetism to a one-dimensional spin liquid},
  author = {Heidarian, Dariush and Sorella, Sandro and Becca, Federico},
  journal = {Phys. Rev. B},
  volume = {80},
  issue = {1},
  pages = {012404},
  numpages = {4},
  year = {2009},
  month = {Jul},
  publisher = {American Physical Society},
  doi = {10.1103/PhysRevB.80.012404},
  url = {https://link.aps.org/doi/10.1103/PhysRevB.80.012404}
}

@article{ABC2024,
  title = {Multiferroicity and Topology in Twisted Transition Metal Dichalcogenides},
  author = {Abouelkomsan, Ahmed and Bergholtz, Emil J. and Chatterjee, Shubhayu},
  journal = {Phys. Rev. Lett.},
  volume = {133},
  issue = {2},
  pages = {026801},
  numpages = {8},
  year = {2024},
  month = {Jul},
  publisher = {American Physical Society},
  doi = {10.1103/PhysRevLett.133.026801},
  url = {https://link.aps.org/doi/10.1103/PhysRevLett.133.026801}
}

@article{pavarini2008,
  title = {Mechanism for Orbital Ordering in ${\mathrm{KCuF}}_{3}$},
  author = {Pavarini, E. and Koch, E. and Lichtenstein, A. I.},
  journal = {Phys. Rev. Lett.},
  volume = {101},
  issue = {26},
  pages = {266405},
  numpages = {4},
  year = {2008},
  month = {Dec},
  publisher = {American Physical Society},
  doi = {10.1103/PhysRevLett.101.266405},
  url = {https://link.aps.org/doi/10.1103/PhysRevLett.101.266405}
}

@article{Koga2018,
  title = {Role of spin-orbit coupling in the Kugel-Khomskii model on the honeycomb lattice},
  author = {Koga, Akihisa and Nakauchi, Shiryu and Nasu, Joji},
  journal = {Phys. Rev. B},
  volume = {97},
  issue = {9},
  pages = {094427},
  numpages = {5},
  year = {2018},
  month = {Mar},
  publisher = {American Physical Society},
  doi = {10.1103/PhysRevB.97.094427},
  url = {https://link.aps.org/doi/10.1103/PhysRevB.97.094427}
}

@article{Iwazaki2023,
  title = {Material-based analysis of spin-orbital Mott insulators},
  author = {Iwazaki, Ryuta and Shinaoka, Hiroshi and Hoshino, Shintaro},
  journal = {Phys. Rev. B},
  volume = {108},
  issue = {24},
  pages = {L241108},
  numpages = {7},
  year = {2023},
  month = {Dec},
  publisher = {American Physical Society},
  doi = {10.1103/PhysRevB.108.L241108},
  url = {https://link.aps.org/doi/10.1103/PhysRevB.108.L241108}
}

@article{Singh2010,
  title = {Antiferromagnetic Mott insulating state in single crystals of the honeycomb lattice material ${\text{Na}}_{2}{\text{IrO}}_{3}$},
  author = {Singh, Yogesh and Gegenwart, P.},
  journal = {Phys. Rev. B},
  volume = {82},
  issue = {6},
  pages = {064412},
  numpages = {7},
  year = {2010},
  month = {Aug},
  publisher = {American Physical Society},
  doi = {10.1103/PhysRevB.82.064412},
  url = {https://link.aps.org/doi/10.1103/PhysRevB.82.064412}
}

@article{Comin2012,
  title = {${\mathrm{Na}}_{2}{\mathrm{IrO}}_{3}$ as a Novel Relativistic Mott Insulator with a 340-meV Gap},
  author = {Comin, R. and Levy, G. and Ludbrook, B. and Zhu, Z.-H. and Veenstra, C. N. and Rosen, J. A. and Singh, Yogesh and Gegenwart, P. and Stricker, D. and Hancock, J. N. and van der Marel, D. and Elfimov, I. S. and Damascelli, A.},
  journal = {Phys. Rev. Lett.},
  volume = {109},
  issue = {26},
  pages = {266406},
  numpages = {5},
  year = {2012},
  month = {Dec},
  publisher = {American Physical Society},
  doi = {10.1103/PhysRevLett.109.266406},
  url = {https://link.aps.org/doi/10.1103/PhysRevLett.109.266406}
}

@article{Li2015,
  title = {Rare-Earth Triangular Lattice Spin Liquid: A Single-Crystal Study of ${\mathrm{YbMgGaO}}_{4}$},
  author = {Li, Yuesheng and Chen, Gang and Tong, Wei and Pi, Li and Liu, Juanjuan and Yang, Zhaorong and Wang, Xiaoqun and Zhang, Qingming},
  journal = {Phys. Rev. Lett.},
  volume = {115},
  issue = {16},
  pages = {167203},
  numpages = {6},
  year = {2015},
  month = {Oct},
  publisher = {American Physical Society},
  doi = {10.1103/PhysRevLett.115.167203},
  url = {https://link.aps.org/doi/10.1103/PhysRevLett.115.167203}
}

@article{Li2016,
  title = {Anisotropic spin model of strong spin-orbit-coupled triangular antiferromagnets},
  author = {Li, Yao-Dong and Wang, Xiaoqun and Chen, Gang},
  journal = {Phys. Rev. B},
  volume = {94},
  issue = {3},
  pages = {035107},
  numpages = {12},
  year = {2016},
  month = {Jul},
  publisher = {American Physical Society},
  doi = {10.1103/PhysRevB.94.035107},
  url = {https://link.aps.org/doi/10.1103/PhysRevB.94.035107}
}

@Article{tenpy2024,
    title={{Tensor network Python (TeNPy) version 1}},
    author={Johannes Hauschild and Jakob Unfried and Sajant Anand and Bartholomew Andrews and Marcus Bintz and Umberto Borla and Stefan Divic and Markus Drescher and Jan Geiger and Martin Hefel and Kévin Hémery and Wilhelm Kadow and Jack Kemp and Nico Kirchner and Vincent S. Liu and Gunnar Möller and Daniel Parker and Michael Rader and Anton Romen and Samuel Scalet and Leon Schoonderwoerd and Maximilian Schulz and Tomohiro Soejima and Philipp Thoma and Yantao Wu and Philip Zechmann and Ludwig Zweng and Roger S. K. Mong and Michael P. Zaletel and Frank Pollmann},
    journal={SciPost Phys. Codebases},
    pages={41},
    year={2024},
    publisher={SciPost},
    doi={10.21468/SciPostPhysCodeb.41},
    url={https://scipost.org/10.21468/SciPostPhysCodeb.41},
}

@article{Zhang2021,
  title = {Electronic structures, charge transfer, and charge order in twisted transition metal dichalcogenide bilayers},
  author = {Zhang, Yang and Liu, Tongtong and Fu, Liang},
  journal = {Phys. Rev. B},
  volume = {103},
  issue = {15},
  pages = {155142},
  numpages = {6},
  year = {2021},
  month = {Apr},
  publisher = {American Physical Society},
  doi = {10.1103/PhysRevB.103.155142},
  url = {https://link.aps.org/doi/10.1103/PhysRevB.103.155142}
}

@book{AuerbachBook,
  title={Interacting electrons and quantum magnetism},
  author={Auerbach, Assa},
  year={2012},
  publisher={Springer Science \& Business Media}
}

@article{Kitaev_2006,
   title={Anyons in an exactly solved model and beyond},
   volume={321},
   ISSN={0003-4916},
   url={http://dx.doi.org/10.1016/j.aop.2005.10.005},
   DOI={10.1016/j.aop.2005.10.005},
   number={1},
   journal={Annals of Physics},
   publisher={Elsevier BV},
   author={Kitaev, Alexei},
   year={2006},
   month=jan, pages={2–111} }

@article{Thouless_1965,
doi = {10.1088/0370-1328/86/5/301},
url = {https://dx.doi.org/10.1088/0370-1328/86/5/301},
year = {1965},
month = {nov},
publisher = {},
volume = {86},
number = {5},
pages = {893},
author = {D J Thouless},
title = {Exchange in solid 3He and the Heisenberg Hamiltonian},
journal = {Proceedings of the Physical Society}
}

@article{Raghu2008,
  title = {Topological Mott Insulators},
  author = {Raghu, S. and Qi, Xiao-Liang and Honerkamp, C. and Zhang, Shou-Cheng},
  journal = {Phys. Rev. Lett.},
  volume = {100},
  issue = {15},
  pages = {156401},
  numpages = {4},
  year = {2008},
  month = {Apr},
  publisher = {American Physical Society},
  doi = {10.1103/PhysRevLett.100.156401},
  url = {https://link.aps.org/doi/10.1103/PhysRevLett.100.156401}
}

@article{Olariu2009,
  title = {Spin dynamics in Heisenberg triangular antiferromagnets: A $\ensuremath{\mu}\text{SR}$ study of ${\text{LiCrO}}_{2}$},
  author = {Olariu, A. and Mendels, P. and Bert, F. and Alexander, L. K. and Mahajan, A. V. and Hillier, A. D. and Amato, A.},
  journal = {Phys. Rev. B},
  volume = {79},
  issue = {22},
  pages = {224401},
  numpages = {5},
  year = {2009},
  month = {Jun},
  publisher = {American Physical Society},
  doi = {10.1103/PhysRevB.79.224401},
  url = {https://link.aps.org/doi/10.1103/PhysRevB.79.224401}
}

@article{Luo2018,
  title = {Spinon Magnetic Resonance of Quantum Spin Liquids},
  author = {Luo, Zhu-Xi and Lake, Ethan and Mei, Jia-Wei and Starykh, Oleg A.},
  journal = {Phys. Rev. Lett.},
  volume = {120},
  issue = {3},
  pages = {037204},
  numpages = {6},
  year = {2018},
  month = {Jan},
  publisher = {American Physical Society},
  doi = {10.1103/PhysRevLett.120.037204},
  url = {https://link.aps.org/doi/10.1103/PhysRevLett.120.037204}
}

@book{gradshteyn2007,
  author = {Gradshteyn, I. S. and Ryzhik, I. M.},
  edition = {Seventh},
  isbn = {978-0-12-373637-6; 0-12-373637-4},
  mrclass = {00A22 (33-00 65-00 65A05)},
  mrnumber = {2360010 (2008g:00005)},
  pages = {xlviii+1171},
  publisher = {Elsevier/Academic Press, Amsterdam},
  title = {Table of integrals, series, and products},
  year = 2007
}

@software{kofod2025,
  title = {{{JuliaNLSolvers}}/{{Optim}}.Jl: V1.13.2 (Docs)},
  shorttitle = {{{JuliaNLSolvers}}/{{Optim}}.Jl},
  author = {Patrick {Kofod Mogensen} and John Myles White and Asbjørn Nilsen Riseth and Tim Holy and Miles Lubin and Christof and Andreas Noack and Antoine Levitt and Benoît Legat and Christoph Ortner and Blake Johnson and Christopher Rackauckas and Yichao Yu and Kristoffer Carlsson and Dahua Lin and Arno Strouwen and Josua Grawitter and Takafumi Arakaki and Benoît Pasquier and {abhro} and Thomas R. Covert and Ron Rock and Michael Creel and {cossio} and Jeffrey Regier and David Widmann and Ben Kuhn and Alexey Stukalov},
  date = {2025-06-12},
  doi = {10.5281/ZENODO.15649599},
  url = {https://zenodo.org/doi/10.5281/zenodo.15649599},
  urldate = {2025-06-27},
  organization = {Zenodo},
  version = {v1.13.2}
}

@article{edelman1998,
  title={The geometry of algorithms with orthogonality constraints},
  author={Edelman, Alan and Arias, Tom{\'a}s A and Smith, Steven T},
  journal={SIAM journal on Matrix Analysis and Applications},
  volume={20},
  number={2},
  pages={303--353},
  year={1998},
  publisher={SIAM}
}

@incollection{absil2008,
  title={Optimization algorithms on matrix manifolds},
  author={Absil, P-A and Mahony, Robert and Sepulchre, Rodolphe},
  booktitle={Optimization Algorithms on Matrix Manifolds},
  year={2009},
  publisher={Princeton University Press}
}

@article{JK2009,
  title = {Mott Insulators in the Strong Spin-Orbit Coupling Limit: From Heisenberg to a Quantum Compass and Kitaev Models},
  author = {Jackeli, G. and Khaliullin, G.},
  journal = {Phys. Rev. Lett.},
  volume = {102},
  issue = {1},
  pages = {017205},
  numpages = {4},
  year = {2009},
  month = {Jan},
  publisher = {American Physical Society},
  doi = {10.1103/PhysRevLett.102.017205},
  url = {https://link.aps.org/doi/10.1103/PhysRevLett.102.017205}
}

@article{CJK2010,
  title = {Kitaev-Heisenberg Model on a Honeycomb Lattice: Possible Exotic Phases in Iridium Oxides ${A}_{2}{\mathrm{IrO}}_{3}$},
  author = {Chaloupka, Ji\ifmmode\check{r}\else\v{r}\fi{}\'{\i} and Jackeli, George and Khaliullin, Giniyat},
  journal = {Phys. Rev. Lett.},
  volume = {105},
  issue = {2},
  pages = {027204},
  numpages = {4},
  year = {2010},
  month = {Jul},
  publisher = {American Physical Society},
  doi = {10.1103/PhysRevLett.105.027204},
  url = {https://link.aps.org/doi/10.1103/PhysRevLett.105.027204}
}

@article{takagi2019concept,
  author = {Takagi, Hidenori and Takayama, Tomohiro and Jackeli, George and Khaliullin, Giniyat and Nagler, Stephen E.},
  title = {Concept and realization of Kitaev quantum spin liquids},
  journal = {Nature Reviews Physics},
  year = {2019},
  volume = {1},
  number = {4},
  pages = {264--280},
  doi = {10.1038/s42254-019-0038-2},
  url = {https://doi.org/10.1038/s42254-019-0038-2}
}

@article{WVZC2024,
  title = {Electrical Control of Spin and Valley in Spin-Orbit Coupled Graphene Multilayers},
  author = {Wang, Taige and Vila, Marc and Zaletel, Michael P. and Chatterjee, Shubhayu},
  journal = {Phys. Rev. Lett.},
  volume = {132},
  issue = {11},
  pages = {116504},
  numpages = {7},
  year = {2024},
  month = {Mar},
  publisher = {American Physical Society},
  doi = {10.1103/PhysRevLett.132.116504},
  url = {https://link.aps.org/doi/10.1103/PhysRevLett.132.116504}
}

@article{Ramesh2021,
author = {Ramesh, Ramamoorthy  and Manipatruni, Sasikanth },
title = {Electric field control of magnetism},
journal = {Proceedings of the Royal Society A: Mathematical, Physical and Engineering Sciences},
volume = {477},
number = {2251},
pages = {20200942},
year = {2021},
doi = {10.1098/rspa.2020.0942}
}

@article{ZKS98,
  title = {Landau theory of stripe phases in cuprates and nickelates},
  author = {Zachar, Oron and Kivelson, S. A. and Emery, V. J.},
  journal = {Phys. Rev. B},
  volume = {57},
  issue = {3},
  pages = {1422--1426},
  numpages = {0},
  year = {1998},
  month = {Jan},
  publisher = {American Physical Society},
  doi = {10.1103/PhysRevB.57.1422},
  url = {https://link.aps.org/doi/10.1103/PhysRevB.57.1422}
}

@Article{Matsukura2015,
author={Matsukura, Fumihiro
and Tokura, Yoshinori
and Ohno, Hideo},
title={Control of magnetism by electric fields},
journal={Nature Nanotechnology},
year={2015},
month={Mar},
day={01},
volume={10},
number={3},
pages={209-220},
issn={1748-3395},
doi={10.1038/nnano.2015.22},
url={https://doi.org/10.1038/nnano.2015.22}
}

@Article{Sierra2021,
author={Sierra, Juan F.
and Fabian, Jaroslav
and Kawakami, Roland K.
and Roche, Stephan
and Valenzuela, Sergio O.},
title={Van der Waals heterostructures for spintronics and opto-spintronics},
journal={Nature Nanotechnology},
year={2021},
month={Aug},
day={01},
volume={16},
number={8},
pages={856-868},
doi={10.1038/s41565-021-00936-x},
url={https://doi.org/10.1038/s41565-021-00936-x}
}

@Article{DGG_2020_TBGswitch,
author={He, Wen-Yu
and Goldhaber-Gordon, David
and Law, K. T.},
title={Giant orbital magnetoelectric effect and current-induced magnetization switching in twisted bilayer graphene},
journal={Nature Communications},
year={2020},
month={Apr},
day={03},
volume={11},
number={1},
pages={1650},
doi={10.1038/s41467-020-15473-9},
url={https://doi.org/10.1038/s41467-020-15473-9}
}

@article{andrei2021marvels,
  author = {Andrei, Eva Y. and Efetov, Dmitri K. and Jarillo-Herrero, Pablo and MacDonald, Allan H. and Mak, Kin Fai and Senthil, T. and Tutuc, Emanuel and Yazdani, Ali and Young, Andrea F.},
  title = {The marvels of moir{\'e} materials},
  journal = {Nature Reviews Materials},
  year = {2021},
  volume = {6},
  number = {3},
  pages = {201--206},
  doi = {10.1038/s41578-021-00284-1},
  url = {https://doi.org/10.1038/s41578-021-00284-1}
}

@article{Polshyn2020,
author={Polshyn, H.
and Zhu, J.
and Kumar, M. A.
and Zhang, Y.
and Yang, F.
and Tschirhart, C. L.
and Serlin, M.
and Watanabe, K.
and Taniguchi, T.
and MacDonald, A. H.
and Young, A. F.},
title={Electrical switching of magnetic order in an orbital Chern insulator},
journal={Nature},
year={2020},
month={Dec},
day={01},
volume={588},
number={7836},
pages={66-70},
issn={1476-4687},
doi={10.1038/s41586-020-2963-8},
url={https://doi.org/10.1038/s41586-020-2963-8}
}

@article{Zhu2020,
  title = {Voltage-Controlled Magnetic Reversal in Orbital Chern Insulators},
  author = {Zhu, Jihang and Su, Jung-Jung and MacDonald, A. H.},
  journal = {Phys. Rev. Lett.},
  volume = {125},
  issue = {22},
  pages = {227702},
  numpages = {6},
  year = {2020},
  month = {Nov},
  publisher = {American Physical Society},
  doi = {10.1103/PhysRevLett.125.227702},
  url = {https://link.aps.org/doi/10.1103/PhysRevLett.125.227702}
}

@article{Naik2020,
  title = {Origin and evolution of ultraflat bands in twisted bilayer transition metal dichalcogenides: Realization of triangular quantum dots},
  author = {Naik, Mit H. and Kundu, Sudipta and Maity, Indrajit and Jain, Manish},
  journal = {Phys. Rev. B},
  volume = {102},
  issue = {7},
  pages = {075413},
  numpages = {11},
  year = {2020},
  month = {Aug},
  publisher = {American Physical Society},
  doi = {10.1103/PhysRevB.102.075413},
  url = {https://link.aps.org/doi/10.1103/PhysRevB.102.075413}
}

@article{Manchon_2015,
   title={New perspectives for Rashba spin–orbit coupling},
   volume={14},
   ISSN={1476-4660},
   url={http://dx.doi.org/10.1038/nmat4360},
   DOI={10.1038/nmat4360},
   number={9},
   journal={Nature Materials},
   publisher={Springer Science and Business Media LLC},
   author={Manchon, A. and Koo, H. C. and Nitta, J. and Frolov, S. M. and Duine, R. A.},
   year={2015},
   month=aug, pages={871–882} }

@article{bihlmayer2022rashba,
  author = {Bihlmayer, Gustav and Noël, Paul and Vyalikh, Denis V. and Chulkov, Evgueni V. and Manchon, Aurélien},
  title = {Rashba-like physics in condensed matter},
  journal = {Nature Reviews Physics},
  year = {2022},
  volume = {4},
  number = {10},
  pages = {642--659},
  doi = {10.1038/s42254-022-00490-y},
  url = {https://doi.org/10.1038/s42254-022-00490-y}
}

@article{shi2023recent,
  author = {Shi, Shilei and Wang, Xiaomin and Zhao, Yan and Zhao, Weisheng},
  title = {Recent progress in strong spin-orbit coupling van der waals materials and their heterostructures for spintronic applications},
  journal = {Materials Today Electronics},
  year = {2023},
  volume = {6},
  pages = {100060},
  doi = {10.1016/j.mtelec.2023.100060},
  url = {https://doi.org/10.1016/j.mtelec.2023.100060}
}

@article{zhang2023enhanced,
  author = {Zhang, Yiran and Polski, Robert and Thomson, Alex and Lantagne-Hurtubise, {\'E}tienne and Lewandowski, Cyprian and Zhou, Haoxin and Watanabe, Kenji and Taniguchi, Takashi and Alicea, Jason and Nadj-Perge, Stevan},
  title = {Enhanced superconductivity in spin-orbit proximitized bilayer graphene},
  journal = {Nature},
  year = {2023},
  volume = {613},
  number = {7943},
  pages = {268--273},
  doi = {10.1038/s41586-022-05446-x},
  url = {https://doi.org/10.1038/s41586-022-05446-x}
}

@article{zhang2025twist,
  author = {Zhang, Yiran and Shavit, Gal and Ma, Huiyang and Han, Youngjoon and Siu, Chi Wang and Mukherjee, Ankan and Watanabe, Kenji and Taniguchi, Takashi and Hsieh, David and Lewandowski, Cyprian and von Oppen, Felix and Oreg, Yuval and Nadj-Perge, Stevan},
  title = {Twist-programmable superconductivity in spin--orbit-coupled bilayer graphene},
  journal = {Nature},
  year = {2025},
  volume = {641},
  number = {8063},
  pages = {625--631},
  doi = {10.1038/s41586-025-08959-3},
  url = {https://doi.org/10.1038/s41586-025-08959-3}
}

@article{patterson2025superconductivity,
  author = {Patterson, Caitlin L. and Sheekey, Owen I. and Arp, Trevor B. and Holleis, Ludwig F. and Koh, Jin Ming and Choi, Youngneak and Xie, Tianji and Xu, Shuyang and Guo, Yiran and Stoyanov, Hristo and Watanabe, Kenji and Taniguchi, Takashi and Alicea, Jason and Lantagne-Hurtubise, {\'E}tienne and Nadj-Perge, Stevan and Young, Andrea F.},
  title = {Superconductivity and spin canting in spin--orbit-coupled trilayer graphene},
  journal = {Nature},
  year = {2025},
  volume = {641},
  number = {8063},
  pages = {632--638},
  doi = {10.1038/s41586-025-08863-w},
  url = {https://doi.org/10.1038/s41586-025-08863-w}
}

@article{mak2022semiconductor,
  author = {Mak, Kin Fai and Shan, Jie},
  title = {Semiconductor moir{\'e} materials},
  journal = {Nature Nanotechnology},
  year = {2022},
  volume = {17},
  number = {7},
  pages = {686--695},
  doi = {10.1038/s41565-022-01165-6},
  url = {https://doi.org/10.1038/s41565-022-01165-6}
}

@article{nuckolls2024microscopic,
  author = {Nuckolls, Kevin P. and Yazdani, Ali},
  title = {A microscopic perspective on moir{\'e} materials},
  journal = {Nature Reviews Materials},
  year = {2024},
  volume = {9},
  number = {7},
  pages = {460--480},
  doi = {10.1038/s41578-024-00682-1},
  url = {https://doi.org/10.1038/s41578-024-00682-1}
}

@book{giamarchi2003quantum,
  title={Quantum physics in one dimension},
  author={Giamarchi, Thierry},
  volume={121},
  year={2003},
  publisher={Clarendon press}
}

@article{QxtoDx2024,
author={Meng, Yuze
and Ma, Lei
and Yan, Li
and Khalifa, Ahmed
and Chen, Dongxue
and Zhang, Shuai
and Banerjee, Rounak
and Taniguchi, Takashi
and Watanabe, Kenji
and Tongay, Seth Ariel
and Hunt, Benjamin
and Lin, Shi-Zeng
and Yao, Wang
and Cui, Yong-Tao
and Chatterjee, Shubhayu
and Shi, Su-Fei},
title={Strong-interaction-driven quadrupolar-to-dipolar exciton transitions in a trilayer moir{\'e} superlattice},
journal={Nature Photonics},
year={2025},
month={Aug},
day={21},
issn={1749-4893},
doi={10.1038/s41566-025-01741-x},
url={https://doi.org/10.1038/s41566-025-01741-x}
}

@article{Troyer2005,
  title = {Two-Step Restoration of SU(2) Symmetry in a Frustrated Ring-Exchange Magnet},
  author = {L\"auchli, A. and Domenge, J. C. and Lhuillier, C. and Sindzingre, P. and Troyer, M.},
  journal = {Phys. Rev. Lett.},
  volume = {95},
  issue = {13},
  pages = {137206},
  numpages = {4},
  year = {2005},
  month = {Sep},
  publisher = {American Physical Society},
  doi = {10.1103/PhysRevLett.95.137206},
  url = {https://link.aps.org/doi/10.1103/PhysRevLett.95.137206}
}

@article{Katsura2005,
  title = {Spin Current and Magnetoelectric Effect in Noncollinear Magnets},
  author = {Katsura, Hosho and Nagaosa, Naoto and Balatsky, Alexander V.},
  journal = {Phys. Rev. Lett.},
  volume = {95},
  issue = {5},
  pages = {057205},
  numpages = {4},
  year = {2005},
  month = {Jul},
  publisher = {American Physical Society},
  doi = {10.1103/PhysRevLett.95.057205},
  url = {https://link.aps.org/doi/10.1103/PhysRevLett.95.057205}
}

@article{xiao2012coupled,
  title = {Coupled Spin and Valley Physics in Monolayers of ${\mathrm{MoS}}_{2}$ and Other Group-VI Dichalcogenides},
  author = {Xiao, Di and Liu, Gui-Bin and Feng, Wanxiang and Xu, Xiaodong and Yao, Wang},
  journal = {Phys. Rev. Lett.},
  volume = {108},
  issue = {19},
  pages = {196802},
  numpages = {5},
  year = {2012},
  month = {May},
  publisher = {American Physical Society},
  doi = {10.1103/PhysRevLett.108.196802},
  url = {https://link.aps.org/doi/10.1103/PhysRevLett.108.196802}
}

@article{sahay2021noise,
  title={Noise electrometry of polar and dielectric materials},
  author={Sahay, Rahul and Hsieh, Satcher and Parsonnet, Eric and Martin, Lane W and Ramesh, Ramamoorthy and Yao, Norman Y and Chatterjee, Shubhayu},
  journal={arXiv preprint arXiv:2111.09315},
  year={2021},
  doi={https://doi.org/10.48550/arXiv.2111.09315},
  url={https://arxiv.org/abs/2111.09315}
}

@article{machado2023quantum,
  title = {Quantum Noise Spectroscopy of Dynamical Critical Phenomena},
  author = {Machado, Francisco and Demler, Eugene A. and Yao, Norman Y. and Chatterjee, Shubhayu},
  journal = {Phys. Rev. Lett.},
  volume = {131},
  issue = {7},
  pages = {070801},
  numpages = {8},
  year = {2023},
  month = {Aug},
  publisher = {American Physical Society},
  doi = {10.1103/PhysRevLett.131.070801},
  url = {https://link.aps.org/doi/10.1103/PhysRevLett.131.070801}
}

@article{CRD2019,
  title = {Diagnosing phases of magnetic insulators via noise magnetometry with spin qubits},
  author = {Chatterjee, Shubhayu and Rodriguez-Nieva, Joaquin F. and Demler, Eugene},
  journal = {Phys. Rev. B},
  volume = {99},
  issue = {10},
  pages = {104425},
  numpages = {23},
  year = {2019},
  month = {Mar},
  publisher = {American Physical Society},
  doi = {10.1103/PhysRevB.99.104425},
  url = {https://link.aps.org/doi/10.1103/PhysRevB.99.104425}
}

@article{SD2024,
  title = {Superconductivity in a topological lattice model with strong repulsion},
  author = {Sahay, Rahul and Divic, Stefan and Parker, Daniel E. and Soejima, Tomohiro and Anand, Sajant and Hauschild, Johannes and Aidelsburger, Monika and Vishwanath, Ashvin and Chatterjee, Shubhayu and Yao, Norman Y. and Zaletel, Michael P.},
  journal = {Phys. Rev. B},
  volume = {110},
  issue = {19},
  pages = {195126},
  numpages = {26},
  year = {2024},
  month = {Nov},
  publisher = {American Physical Society},
  doi = {10.1103/PhysRevB.110.195126},
  url = {https://link.aps.org/doi/10.1103/PhysRevB.110.195126}
}

@article{Foutty2023,
author={Foutty, Benjamin A.
and Yu, Jiachen
and Devakul, Trithep
and Kometter, Carlos R.
and Zhang, Yang
and Watanabe, Kenji
and Taniguchi, Takashi
and Fu, Liang
and Feldman, Benjamin E.},
title={Tunable spin and valley excitations of correlated insulators in $\Gamma$-valley moir{\'e} bands},
journal={Nature Materials},
year={2023},
month={Jun},
day={01},
volume={22},
number={6},
pages={731-736},
issn={1476-4660},
doi={10.1038/s41563-023-01534-z},
url={https://doi.org/10.1038/s41563-023-01534-z}
}

@article{KaneMele,
  title = {Quantum Spin Hall Effect in Graphene},
  author = {Kane, C. L. and Mele, E. J.},
  journal = {Phys. Rev. Lett.},
  volume = {95},
  issue = {22},
  pages = {226801},
  numpages = {4},
  year = {2005},
  month = {Nov},
  publisher = {American Physical Society},
  doi = {10.1103/PhysRevLett.95.226801},
  url = {https://link.aps.org/doi/10.1103/PhysRevLett.95.226801}
}

@article{tschirhart2021imaging,
  author = {Tschirhart, C. L. and Serlin, M. and Polshyn, H. and Shragai, A. and Xia, Z. and Zhu, J. and Zhang, Y. and Watanabe, K. and Taniguchi, T. and Huber, M. E. and Young, A. F.},
  title = {Imaging orbital ferromagnetism in a moir{\'e} Chern insulator},
  journal = {Science},
  year = {2021},
  volume = {372},
  number = {6548},
  pages = {1323--1327},
  doi = {10.1126/science.abd3190},
  url = {https://doi.org/10.1126/science.abd3190}
}

@article{anderson2023programming,
  author = {Anderson, Eric and Fan, Feng-Ren and Cai, Jiaqi and Holtzmann, William and Taniguchi, Takashi and Watanabe, Kenji and Xiao, Di and Yao, Wang and Xu, Xiaodong},
  title = {Programming correlated magnetic states with gate-controlled moir{\'e} geometry},
  journal = {Science},
  year = {2023},
  volume = {381},
  number = {6655},
  pages = {325--330},
  doi = {10.1126/science.adg4268},
  url = {https://doi.org/10.1126/science.adg4268}
}

@ARTICLE{XY2025,
       author = {{Li}, Weijie and {Redekop}, Evgeny and {Beach}, Christiano Wang and {Zhang}, Canxun and {Zhang}, Xiaowei and {Liu}, Xiaoyu and {Holtzmann}, Will and {Hu}, Chaowei and {Anderson}, Eric and {Park}, Heonjoon and {Taniguchi}, Takashi and {Watanabe}, Kenji and {Chu}, Jiun-haw and {Fu}, Liang and {Cao}, Ting and {Xiao}, Di and {Young}, Andrea F. and {Xu}, Xiaodong},
        title = "{Universal Magnetic Phases in Twisted Bilayer MoTe$_2$}",
      journal = {arXiv e-prints},
         year = 2025,
        month = jul,
          eid = {arXiv:2507.22354},
        pages = {arXiv:2507.22354},
          doi = {10.48550/arXiv.2507.22354},
archivePrefix = {arXiv},
       eprint = {2507.22354},
 primaryClass = {cond-mat.mes-hall},
       adsurl = {https://ui.adsabs.harvard.edu/abs/2025arXiv250722354L}
}

@article{ziffer2024quantum,
  title={Quantum noise spectroscopy of criticality in an atomically thin magnet},
  author={Ziffer, Mark E and Machado, Francisco and Ursprung, Benedikt and Lozovoi, Artur and Tazi, Aya Batoul and Yuan, Zhiyang and Ziebel, Michael E and Delord, Tom and Zeng, Nanyu and Telford, Evan and others},
  journal={arXiv preprint arXiv:2407.05614},
  doi={https://doi.org/10.48550/arXiv.2407.05614},
  year={2024}
}

@article{Joaquin2018,
  title = {Probing one-dimensional systems via noise magnetometry with single spin qubits},
  author = {Rodriguez-Nieva, Joaquin F. and Agarwal, Kartiek and Giamarchi, Thierry and Halperin, Bertrand I. and Lukin, Mikhail D. and Demler, Eugene},
  journal = {Phys. Rev. B},
  volume = {98},
  issue = {19},
  pages = {195433},
  numpages = {16},
  year = {2018},
  month = {Nov},
  publisher = {American Physical Society},
  doi = {10.1103/PhysRevB.98.195433},
  url = {https://link.aps.org/doi/10.1103/PhysRevB.98.195433}
}

@article{rovny2024nanoscale,
  author = {Rovny, Jared and Gopalakrishnan, Sarang and Jayich, Ania C. Bleszynski and Maletinsky, Patrick and Demler, Eugene and de Leon, Nathalie P.},
  title = {Nanoscale diamond quantum sensors for many-body physics},
  journal = {Nature Reviews Physics},
  year = {2024},
  volume = {6},
  number = {12},
  pages = {753--768},
  doi = {10.1038/s42254-024-00775-4},
  url = {https://doi.org/10.1038/s42254-024-00775-4}
}

@article{inbar2023quantum,
  author = {Inbar, Alon and Birkbeck, John and Xiao, Jiacheng and Taniguchi, Takashi and Watanabe, Kenji and Yan, Binghai and Oreg, Yuval and Stern, Ady and Berg, Erez and Ilani, Shahal},
  title = {The quantum twisting microscope},
  journal = {Nature},
  year = {2023},
  volume = {614},
  number = {7949},
  pages = {682--687},
  doi = {10.1038/s41586-022-05685-y},
  url = {https://doi.org/10.1038/s41586-022-05685-y}
}

@article{sidler2017fermi,
  author = {Sidler, Meinrad and Back, Patrick and Cotlet, Ovidiu and Srivastava, Ajit and Fink, Thomas and Kroner, Martin and Demler, Eugene and Imamoglu, Atac},
  title = {Fermi polaron-polaritons in charge-tunable atomically thin semiconductors},
  journal = {Nature Physics},
  year = {2017},
  volume = {13},
  number = {3},
  pages = {255--261},
  doi = {10.1038/nphys3949},
  url = {https://doi.org/10.1038/nphys3949}
}

@article{ImamogluPRL2019,
  title = {Interaction-Induced Shubnikov--de Haas Oscillations in Optical Conductivity of Monolayer ${\mathrm{MoSe}}_{2}$},
  author = {Smole\ifmmode \acute{n}\else \'{n}\fi{}ski, T. and Cotlet, O. and Popert, A. and Back, P. and Shimazaki, Y. and Kn\"uppel, P. and Dietler, N. and Taniguchi, T. and Watanabe, K. and Kroner, M. and Imamoglu, A.},
  journal = {Phys. Rev. Lett.},
  volume = {123},
  issue = {9},
  pages = {097403},
  numpages = {6},
  year = {2019},
  month = {Aug},
  publisher = {American Physical Society},
  doi = {10.1103/PhysRevLett.123.097403},
  url = {https://link.aps.org/doi/10.1103/PhysRevLett.123.097403}
}

@article{KIVC,
  title = {Ground State and Hidden Symmetry of Magic-Angle Graphene at Even Integer Filling},
  author = {Bultinck, Nick and Khalaf, Eslam and Liu, Shang and Chatterjee, Shubhayu and Vishwanath, Ashvin and Zaletel, Michael P.},
  journal = {Phys. Rev. X},
  volume = {10},
  issue = {3},
  pages = {031034},
  numpages = {13},
  year = {2020},
  month = {Aug},
  publisher = {American Physical Society},
  doi = {10.1103/PhysRevX.10.031034},
  url = {https://link.aps.org/doi/10.1103/PhysRevX.10.031034}
}

@article{KIVC2,
  title = {Twisted bilayer graphene. IV. Exact insulator ground states and phase diagram},
  author = {Lian, Biao and Song, Zhi-Da and Regnault, Nicolas and Efetov, Dmitri K. and Yazdani, Ali and Bernevig, B. Andrei},
  journal = {Phys. Rev. B},
  volume = {103},
  issue = {20},
  pages = {205414},
  numpages = {41},
  year = {2021},
  month = {May},
  publisher = {American Physical Society},
  doi = {10.1103/PhysRevB.103.205414},
  url = {https://link.aps.org/doi/10.1103/PhysRevB.103.205414}
}

@article{IKS,
  title = {Kekul\'e Spiral Order at All Nonzero Integer Fillings in Twisted Bilayer Graphene},
  author = {Kwan, Y. H. and Wagner, G. and Soejima, T. and Zaletel, M. P. and Simon, S. H. and Parameswaran, S. A. and Bultinck, N.},
  journal = {Phys. Rev. X},
  volume = {11},
  issue = {4},
  pages = {041063},
  numpages = {23},
  year = {2021},
  month = {Dec},
  publisher = {American Physical Society},
  doi = {10.1103/PhysRevX.11.041063},
  url = {https://link.aps.org/doi/10.1103/PhysRevX.11.041063}
}

@inproceedings{access,
author = {Boerner, Timothy J. and Deems, Stephen and Furlani, Thomas R. and Knuth, Shelley L. and Towns, John},
title = {ACCESS: Advancing Innovation: NSF’s Advanced Cyberinfrastructure Coordination Ecosystem: Services \& Support},
year = {2023},
isbn = {9781450399852},
publisher = {Association for Computing Machinery},
address = {New York, NY, USA},
url = {https://doi.org/10.1145/3569951.3597559},
doi = {10.1145/3569951.3597559},
booktitle = {Practice and Experience in Advanced Research Computing 2023: Computing for the Common Good},
pages = {173–176},
numpages = {4},
location = {Portland, OR, USA},
series = {PEARC '23}
}

@article{li:2008a,
  title = {Entanglement {{Spectrum}} as a {{Generalization}} of {{Entanglement Entropy}}: {{Identification}} of {{Topological Order}} in {{Non-Abelian Fractional Quantum Hall Effect States}}},
  shorttitle = {Entanglement {{Spectrum}} as a {{Generalization}} of {{Entanglement Entropy}}},
  author = {Li, Hui and Haldane, F. D. M.},
  year = {2008},
  month = jul,
  journal = {Physical Review Letters},
  volume = {101},
  number = {1},
  pages = {010504},
  publisher = {American Physical Society},
  doi = {10.1103/PhysRevLett.101.010504}
}

@article{qi:2012a,
  title = {General {{Relationship Between}} the {{Entanglement Spectrum}} and the {{Edge State Spectrum}} of {{Topological Quantum States}}},
  author = {Qi, Xiao-Liang and Katsura, Hosho and Ludwig, Andreas W. W.},
  year = {2012},
  month = may,
  journal = {Physical Review Letters},
  volume = {108},
  number = {19},
  eprint = {1103.5437},
  primaryclass = {cond-mat},
  pages = {196402},
  issn = {0031-9007, 1079-7114},
  doi = {10.1103/PhysRevLett.108.196402},
  archiveprefix = {arXiv}
}

@article{Hohenberg,
  title = {Existence of Long-Range Order in One and Two Dimensions},
  author = {Hohenberg, P. C.},
  journal = {Phys. Rev.},
  volume = {158},
  issue = {2},
  pages = {383--386},
  numpages = {0},
  year = {1967},
  month = {Jun},
  publisher = {American Physical Society},
  doi = {10.1103/PhysRev.158.383},
  url = {https://link.aps.org/doi/10.1103/PhysRev.158.383}
}

@article{MerminWagner,
  title = {Absence of Ferromagnetism or Antiferromagnetism in One- or Two-Dimensional Isotropic Heisenberg Models},
  author = {Mermin, N. D. and Wagner, H.},
  journal = {Phys. Rev. Lett.},
  volume = {17},
  issue = {22},
  pages = {1133--1136},
  numpages = {0},
  year = {1966},
  month = {Nov},
  publisher = {American Physical Society},
  doi = {10.1103/PhysRevLett.17.1133},
  url = {https://link.aps.org/doi/10.1103/PhysRevLett.17.1133}
}

@misc{LUS, 
note = {In practice, we allow for unit cells up to size 12. In most cases, we obtain better energetics for smaller unit cells.} }

@misc{PinningField,
note = {Due to inherent constraints on DMRG, we need to add a small pinning field to see this order in correlators of the spin-current, see Appendix~\ref{app:hsl}.}
}

@misc{boundcharge,
note= {{Recall that the (bound) charge density is given by the divergence of the electrical polarization as $\rho_b = - \bm{\nabla} \cdot \bm{P}$}}}

\end{document}